\documentclass[a4paper,10pt,twoside]{cpc-hepnp}

\usepackage{multicol}
\usepackage{graphicx}
\usepackage{booktabs}
\usepackage{amssymb,bm,mathrsfs,bbm,amscd}
\usepackage[tbtags]{amsmath}
\usepackage{lastpage}

\begin{document}

\fancyhead[co]{\footnotesize L. Tiator et al: Baryon Resonance Analysis from
MAID}


\title{Baryon Resonance Analysis from MAID\thanks{Supported by the Deutsche Forschungsgemeinschaft through the
SFB~443 and  by the joint Russian-German Heisenberg-Landau program.}}

\author{%
      L. Tiator$^{1)}$\email{tiator@kph.uni-mainz.de}
\quad D. Drechsel$^{1)}$\quad S.S. Kamalov$^{2)}$  \quad M.
Vanderhaeghen$^{1)}$ } \maketitle

\address{%
1~(Institut f\"ur Kernphysik, Universit\"at Mainz, D-55099 Mainz,
Germany)\\
2~(Bogoliubov Laboratory for Theoretical Physics, JINR, Dubna,
141980 Moscow Region, Russia)\\
}

\begin{abstract}
The unitary isobar model MAID2007 has been used to analyze the recent data of
pion electroproduction. The model contains all four-star resonances in the
region below $W=2$~GeV and both single-$Q^2$ and $Q^2$ dependent transition
form factors could be obtained for the Delta, Roper, $D_{13}(1520)$,
$S_{11}(1535)$, $S_{31}(1620)$, $S_{11}(1650)$, $D_{15}(1675)$, $F_{15}(1680)$
and $P_{13}(1720)$. From the complete world data base, including also $\pi^-$
data on the neutron, also $Q^2$ dependent neutron form factors are obtained.
For all transition form factors we also give convenient numerical
parameterizations that can be used in other reactions. Furthermore, we show how
the transition form factors can be used to obtain empirical transverse charge
densities and our first results are given for the Roper, the S11 and D13
resonances.
\end{abstract}



\begin{multicols}{2}

\section{Introduction}
Our knowledge about the excitation spectrum of the nucleon was originally
provided by elastic pion-nucleon scattering. All the resonances listed in the
Particle Data Tables\cite{PDG08} have been identified by partial-wave analyses
of this process with both Breit-Wigner and pole extraction techniques. From
such analyses we know the resonance masses, widths, and branching ratios into
the $\pi N$ and $\pi\pi N$ channels. These are reliable parameters for the
four-star resonances, with only few exceptions. In particular, there remains
some doubt about the structure of two prominent resonances, the Roper
$P_{11}(1440)$, which appears unusually broad, and the $S_{11}(1535)$, where
the pole can not be uniquely determined, because it lies close to the $\eta N$
threshold.

On the basis of these relatively firm grounds, additional information can be
obtained for the electromagnetic (e.m.) $\gamma N N^*$ couplings through pion
photo- and electroproduction. These couplings are described by electric,
magnetic, and charge transition form factors (FFs), $G_E^*(Q^2)$, $G_M^*(Q^2)$,
and $G_C^*(Q^2)$, or by linear combinations thereof as helicity amplitudes
$A_{1/2}(Q^2)$, $A_{3/2}(Q^2)$, and $S_{1/2}(Q^2)$. So far we have some
reasonable knowledge of the transverse amplitudes $A_{1/2}$ and $A_{3/2}$ at
the real photon point, which are tabulated in the Particle Data Tables. For
finite $Q^2$ the information found in the literature is scarce and until
recently practically nonexistent for the longitudinal amplitudes $S_{1/2}$.

A big step forward was done during the last decade by the experiments at JLab,
where electroproduction of $\pi^0$ and $\pi^+$ have been measured on the
proton. Most of these experiments did not use polarization degrees of freedom,
but the virtual photon in electroproduction always carries longitudinal and
transverse polarizations which are accessible in experiments with large
azimuthal angle coverage. In addition, also some experiments, especially in the
$\Delta(1232)$ region were performed with polarized electrons, polarized target
and even an almost complete experiment was done in Hall A with 16 unpolarized
and recoil polarization observables at $Q^2=1.0$~GeV$^2$.

With our unitary isobar model MAID we have analyzed the electroproduction data
and have obtained transition form factors for all 13 four-star resonances below
$W=2$~GeV. For the proton target in most cases we could obtain both
single-$Q^2$ and $Q^2$-dependent transition form factors, for the neutron
target we parameterized the $Q^2$ dependence in a simpler form as far as the
existing data from the world data base allows.

Furthermore, the precise e.m. FF data, extracted from experiment, allow us to
map out the quark charge densities in a baryon. It was shown possible to define
a proper density interpretation of the form factor data by viewing the baryon
in a light-front frame. This yields information on the spatial distribution of
the quark charge in the plane transverse to the line-of-sight. In this way, the
quark transverse charge densities were mapped out in the
nucleon\cite{Miller:2007uy,Carlson:2007xd}, and in the
deuteron\cite{Carlson:2008zc} based on empirical FF data. To understand the
e.m. structure of a nucleon resonance, it is of interest to use the precise
transition FF data to reveal the spatial distribution of the quark charges that
induce such a transition. In this way, using the empirical information on the
$N \to N^*$ transition form factors from the MAID analysis\cite{MAID07}, the $N
\to \Delta(1232)$ transition charge densities have been mapped out in
Ref.\citep{Carlson:2007xd} and the $N \to N^*(1440)$ in Ref.\citep{Tiator2009}.
In this work, we will extend this method to map out the quark transition charge
densities inducing the $N \to S_{11}(1535)$ and $N \to D_{13}(1520)$ e.m.
excitations.

\section{The MAID ansatz}

In the spirit of a dynamical approach to pion photo- and electroproduction, the
T-matrix of the unitary isobar model is set up with the following ansatz
\begin{equation}
t_{\gamma\pi}(W)=t_{\gamma\pi}^B(W) + t_{\gamma\pi}^{R}(W)\label{eq:1}
\end{equation}
of a background and a resonance T-matrix, where each of them is individually
unitary. This is a very important starting point that will allow us later to
clearly separate resonance and background amplitudes within a Breit-Wigner
concept.

For a specific partial wave the background T-matrix is set-up by a potential
multiplied by pion nucleon scattering amplitudes in accordance with the
K-matrix approximation,
\begin{equation}
 t^{B,\alpha}_{\gamma\pi}(W,Q^2)=v^{B,\alpha}_{\gamma\pi}(W,Q^2)\,[1+it_{\pi N}^{\alpha}(W) ]\, ,
\label{eq:2}
\end{equation}
where only the on-shell part of the pion nucleon rescattering is maintained and
the off-shell part from pion loop contributions is neglected. At threshold it
is well known that this is a bad approximation for $\gamma,\pi^0$ production,
however in the resonance region it is well justified as the main contribution
from pion loop effects is absorbed by the nucleon resonance dressing.

The background  potential $v_{\gamma\pi}^{B,\alpha}(W,Q^2)$ is described by
Born terms obtained with an energy dependent mixing of
pseudovector-pseudoscalar $\pi NN$ coupling and t-channel vector meson
exchanges. The mixing para\-meters and coupling constants were determined from
an analysis of nonresonant multipoles in the appropriate energy regions. In the
latest version, MAID2007, the $S$, $P$, $D$ and $F$ waves of the background
contributions are unitarized as explained above, where the pion-nucleon elastic
scattering amplitudes, $t^{\alpha}_{\pi N}=[\eta_{\alpha}
\exp(2i\delta_{\alpha})-1]/2i$, are described by the phase shifts
$\delta_{\alpha}$ and the inelasticity parameters $\eta_{\alpha}$ taken from
the GWU/SAID analysis\cite{VPI}.

For the resonance contributions we follow Ref.\citep{Maid} and assume
Breit-Wigner forms for the resonance shape,
\begin{eqnarray}
\lefteqn{t_{\gamma\pi}^{R,\alpha}(W,Q^2) = }\nonumber\\
&& {\bar{\cal A}}_{\alpha}^R(W,Q^2)\, \frac{f_{\gamma
N}(W)\Gamma_{tot}\,M_R\,f_{\pi N}(W)}{M_R^2-W^2-iM_R\, \Gamma_{tot}}
\,e^{i\phi_R}\,, \label{eq:3}
\end{eqnarray}
where $f_{\pi N}$ is the usual Breit-Wigner factor describing the decay of a
resonance $R$ with total width $\Gamma_{tot}(W)$ and physical mass $M_R$. The
expressions for $f_{\gamma N}, \, f_{\pi N}$ and $\Gamma_{tot}$ are given in
Ref.\citep{Maid}. The phase $\phi_R(W)$ in Eq.~(\ref{eq:3}) is introduced to
adjust the total phase such that the Fermi-Watson theorem is fulfilled below
two-pion threshold. For the $S$- and $P$-wave multipoles we extend this
unitarization procedure up to $W=1400$ MeV. Because of a lack of further
information, we assume that the phases $\phi_R$ are constant at the higher
energies. In particular we note that the phase $\phi_R$ for the $P_{33}(1232)$
excitation vanishes at $W=M_R=1232$~MeV for all values of $Q^2$. For this
multipole we may even apply the Fermi-Watson theorem up to $W \approx 1600$~MeV
because the inelasticity parameter $\eta_{\alpha}$ remains close to 1. For the
$D$- and $F$-wave resonances, the phases $\phi_R$ are assumed to be constant
and determined from the best fit.

While in the original version of MAID only the 7 most important nucleon
resonances were included with mostly only transverse e.m. couplings, in our new
version all 13 four-star resonances below $W=2$~GeV are included. These are
$P_{33}(1232)$, $P_{11}(1440)$, $D_{13}(1520)$, $S_{11}(1535)$, $S_{31}(1620)$,
$S_{11}(1650)$, $D_{15}(1675)$, $F_{15}(1680)$, $D_{33}(1700)$, $P_{13}(1720)$,
$F_{35}(1905)$, $P_{31}(1910)$ and $F_{37}(1950)$.

\section{Transition form factors}

The resonance couplings $\bar{\mathcal A}_{\alpha}^R(W,Q^2)$ in most cases are
independent of the total energy and depend only on $Q^2$. A typical energy
dependence occurs in MAID2007 e.g. for the $\Delta(1232)$ resonance in terms of
the virtual photon three-momentum $k(W,Q^2)$. For all other resonances which
are discussed here, however, we can assume a simple $Q^2$ dependence,
$\bar{\mathcal A}_{\alpha}(Q^2)$. They can be taken as constants in a
single-Q$^2$ analysis, e.g. in photoproduction, where $Q^2=0$ but also at any
fixed $Q^2$, where enough data with W and $\theta$ variation is available, see
Table~\ref{tab:database}. Alternatively they can also be parameterized as
functions of $Q^2$ in an ansatz like
\begin{equation}
\bar{\mathcal A}_{\alpha}(Q^2) =\bar{\mathcal A}_{\alpha}(0) (1+a_1 Q^2+a_2 Q^4
+a_3 Q^8)\, e^{-b_1 Q^2}\,. \label{eq:ffpar}
\end{equation}
With such an ansatz it is possible to determine the parameters $\bar{\mathcal
A}_{\alpha}(0)$ from a fit to the world database of photoproduction, while the
parameters $a_i$ and $b_1$ can be obtained from a combined fitting of all
electroproduction data at different $Q^2$. The latter procedure we call the
`superglobal fit'. In MAID the photon couplings $\bar{\mathcal A}_{\alpha}$ are
direct input parameters. They are directly related to the helicity couplings
$A_{1/2}, A_{3/2}$ and $S_{1/2}$ of nucleon resonance excitation. For further
details see Ref.\citep{MAID07}.

In Tables~\ref{tab:param-p1},\ref{tab:param-p2} and \ref{tab:param-n} we give
the numerical values of the parameters for our $Q^2$ dependent, `superglobal
fits'. Our parametrization of the $\Delta(1232)$ form factors are more
complicated, in particular due to build-in requirements from low energy
theorems in the long wavelength limit, details are discussed in
Ref.\citep{MAID07}.
\begin{center}
\tabcaption{ \label{tab:database}  Database of pion electroproduction for
energies above the $\Delta$ resonance up to $W=1.7$~GeV, used in our
single-$Q^2$ transition form factor analysis.} \vspace{-0mm} \footnotesize
\begin{tabular*}{80mm}{@{\extracolsep{\fill}}lccc}
\toprule Reference & year & reaction & $Q^2\;(GeV^2)$ \\
\hline
Joo et al.\cite{Joo02} & 2002         & $p\pi^0$ & $ 0.4 - 1.8 $ \\
Joo et al.\cite{Joo04} & 2004         & $n\pi^+$ & $ 0.4 - 0.65 $ \\
Laveissiere et al.\cite{Lav04} & 2004 & $p\pi^0$ & $ 1.0 $ \\
Egiyan et al.\cite{Egi06} & 2006      & $n\pi^+$ & $ 0.3 - 0.6 $ \\
Ungaro et al.\cite{Ung06} & 2006      & $p\pi^0$ & $ 3.0 - 6.0 $ \\
Park et al.\cite{Par08}   & 2008      & $n\pi^+$ & $ 1.7 - 4.5 $ \\
\bottomrule
\end{tabular*}%
\end{center}
For all other resonances the parameters are listed in the three tables. Due to
the 2008 $\pi^+$ data that have been recently included in our database, we find
differences compared to our MAID2007 parametrization for the following 6 proton
transition form factors to $P_{11}(1440)$, $D_{13}(1520)$, $D_{33}(1440)$,
$D_{15}(1675)$, $F_{15}(1680)$ and $P_{13}(1720)$.

\end{multicols}

\begin{center}
\tabcaption{ \label{tab:param-p1} New parameterizations of our transition form
factors, Eq.~(\ref{eq:ffpar}), for proton targets.} \vspace{-3mm} \footnotesize
\begin{tabular*}{170mm}{@{\extracolsep{\fill}}ccccccc}
\toprule $N^*,\;\Delta^*$ & & $\bar{\mathcal A}_{\alpha}(0)\;(10^{-3}\,GeV^{-1/2})$ &
$a_1\;(GeV^{-2})$ & $a_2\;(GeV^{-4})$ & $a_3\;(GeV^{-8})$ & $b_1\;(GeV^{-2})$ \\
\hline
$P_{11}(1440)p$ & $A_{1/2}$ & $-61.4$ & $ 0.871$ & $-3.516$ & $-0.158$ & $1.36$\\
                & $S_{1/2}$ & $  4.2$ & $  40. $ & $  0   $ & $ 1.50 $ & $1.75$\\
\hline
$D_{13}(1520)p$ & $A_{1/2}$ & $-27.4$ & $ 8.580$ & $-0.252$ & $ 0.357$ & $1.20$\\
                & $A_{3/2}$ & $160.6$ & $-0.820$ & $ 0.541$ & $-0.016$ & $1.06$\\
                & $S_{1/2}$ & $-63.5$ & $ 4.19 $ & $  0   $ & $   0  $ & $3.40$\\
\hline
$D_{15}(1675)p$ & $A_{1/2}$ & $ 15.3$ & $ 0.10 $ & $  0   $ & $   0  $ & $2.00$\\
                & $A_{3/2}$ & $ 21.6$ & $ 1.91 $ & $ 0.18 $ & $   0  $ & $0.69$\\
                & $S_{1/2}$ & $  1.1$ & $  0   $ & $  0   $ & $   0  $ & $2.00$\\
\hline
$F_{15}(1680)p$ & $A_{1/2}$ & $-25.1$ & $ 3.780$ & $-0.292$ & $ 0.080$ & $1.25$\\
                & $A_{3/2}$ & $134.3$ & $ 1.016$ & $ 0.222$ & $ 0.237$ & $2.41$\\
                & $S_{1/2}$ & $-44.0$ & $ 3.783$ & $  0   $ & $   0  $ & $1.85$\\
\hline
$D_{33}(1700) $ & $A_{1/2}$ & $226. $ & $ 1.91 $ & $  0   $ & $   0  $ & $1.77$\\
                & $A_{3/2}$ & $210. $ & $ 0.88 $ & $ 1.71 $ & $   0  $ & $2.02$\\
                & $S_{1/2}$ & $  2.1$ & $   0  $ & $  0   $ & $   0  $ & $2.00$\\
\hline
$P_{13}(1720)p$ & $A_{1/2}$ & $ 73.0$ & $ 1.89 $ & $  0   $ & $   0  $ & $1.55$\\
                & $A_{3/2}$ & $-11.5$ & $10.83 $ & $-0.66 $ & $   0  $ & $0.43$\\
                & $S_{1/2}$ & $-53.0$ & $ 2.46 $ & $  0   $ & $   0  $ & $1.55$\\
\bottomrule
\end{tabular*}%
\end{center}

\begin{multicols}{2}

Above the third resonance region there is an energy gap between
$1800-1900$~MeV, where no four-star resonance can be found. Beyond this gap and
up to 2~GeV three more four-star resonances, $F_{35}(1905)$, $P_{31}(1910)$ and
$F_{37}(1950)$ are reported by the PDG, which are also included in MAID. In
electroproduction nothing is practically known about these states and we have
just introduced their reported photon couplings, multiplied with a simple
gaussian form factor, $exp(-2.0\,Q^2/GeV^2)$. In MAID their main role is to
define a global high-energy behavior that is needed for applications with
dispersion relations and sum rules. Future experiments in this region will give
us the necessary information to map out these form factors in more details.\\

\newpage
\begin{center}
\tabcaption{ \label{tab:param-p2} Maid2007 parameterizations,
Eq.~(\ref{eq:ffpar}), for proton targets ($a_2=a_3=0)$.}
\vspace{-0mm}
\footnotesize
\begin{tabular*}{80mm}{@{\extracolsep{\fill}}ccccc}
\toprule $N^*,\;\Delta^*$ & & $\bar{\mathcal A}_{\alpha}(0)$ & $a_1$ & $b_1$ \\
\hline
$S_{11}(1535)p$ & $A_{1/2}$ & $ 66.4$ & $ 1.608$ & $0.70$\\
                & $S_{1/2}$ & $ -2.0$ & $  23.9$ & $0.81$\\
\hline
$S_{31}(1620) $ & $A_{1/2}$ & $ 65.6$ & $ 1.86 $ & $2.50$\\
                & $S_{1/2}$ & $ 16.2$ & $ 2.83 $ & $2.00$\\
\hline
$S_{11}(1650)p$ & $A_{1/2}$ & $ 33.3$ & $ 1.45 $ & $0.62$\\
                & $S_{1/2}$ & $ -3.5$ & $ 2.88 $ & $0.76$\\
\bottomrule
\end{tabular*}%
\end{center}
\vspace*{0.5cm}

\begin{center}
\tabcaption{ \label{tab:param-n} Same as Table~\ref{tab:param-p2} for neutron
targets.} \vspace{-0mm} \footnotesize
\begin{tabular*}{80mm}{@{\extracolsep{\fill}}ccccc}
\toprule $N^*$ & & $\bar{\mathcal A}_{\alpha}(0)$ & $a_1$ & $b_1$ \\
\hline
$P_{11}(1440)n$ & $A_{1/2}$ & $ 54.1$ & $ 0.95 $ & $1.77$\\
                & $S_{1/2}$ & $-41.5$ & $ 2.98 $ & $1.55$\\
$D_{13}(1520)n$ & $A_{1/2}$ & $-76.5$ & $-0.53 $ & $1.55$\\
                & $A_{3/2}$ & $-154.$ & $ 0.58 $ & $1.75$\\
                & $S_{1/2}$ & $ 13.6$ & $ 15.7 $ & $1.57$\\
$S_{11}(1535)n$ & $A_{1/2}$ & $-50.7$ & $ 4.75 $ & $1.69$\\
                & $S_{1/2}$ & $ 28.5$ & $ 0.36 $ & $1.55$\\
$S_{11}(1650)n$ & $A_{1/2}$ & $  9.3$ & $ 0.13 $ & $1.55$\\
                & $S_{1/2}$ & $ 10. $ & $-0.50 $ & $1.55$\\
$D_{15}(1675)n$ & $A_{1/2}$ & $-61.7$ & $ 0.01 $ & $2.00$\\
                & $A_{3/2}$ & $-83.7$ & $ 0.01 $ & $2.00$\\
                & $S_{1/2}$ & $  0  $ & $  0   $ & $ 0  $\\
$F_{15}(1680)n$ & $A_{1/2}$ & $ 27.9$ & $  0   $ & $1.20$\\
                & $A_{3/2}$ & $-38.4$ & $ 4.09 $ & $1.75$\\
                & $S_{1/2}$ & $  0  $ & $  0   $ & $  0 $\\
$P_{13}(1720)n$ & $A_{1/2}$ & $ -2.9$ & $ 12.70$ & $1.55$\\
                & $A_{3/2}$ & $-31.0$ & $ 5.00 $ & $1.55$\\
                & $S_{1/2}$ & $  0  $ & $   0  $ & $  0 $\\
\bottomrule
\end{tabular*}%
\end{center}

\vspace*{1cm}
\subsection{First Resonance Region}

The $\Delta(1232)P_{33}$ is the only nucleon resonance with a well-defined
Breit-Wigner resonance position, $M_R=1232$~MeV, because it is an ideal
single-channel resonance, where the Breit-Wigner position is identical to the
K-matrix pole position. Therefore, and due to the Watson theorem, the Nucleon
to $\Delta(1232)$ transition is the only case, where we can obtain the form
factors in a practically model independent way.

Results for the $\Delta(1232)$ transitions have been discussed very often in
recent years. The magnetic form factor is very well known up to high momentum
transfer of $Q^2=10$~GeV$^2$ and can be parameterized in a surprisingly simple
form
\begin{equation}
G_M^*(Q^2)=3\,G_D(Q^2) e^{-0.21 Q^2/GeV^2}
\end{equation}
with the standard dipole form factor $G_D$. The electric and Coulomb form
factors are much smaller and are usually given as ratios to the magnetic form
factor. While there is agreement between different analyses on the $E/M$ ratio,
which is practically constant at a few percent with a negative sign the $S/M$
ratio is also negative, but reaches large magnitudes of around $25\%$ at the
$Q^2\approx 6$~GeV$^2$ in the JLab analysis, whereas in the MAID analysis the
magnitude is only around $10\%$ with an asymptotically almost zero slope as
predicted in calculations by Buchmann\cite{Buc04} and Ji et al.\cite{JiM03}.

\subsection{Second Resonance Region}
Above the two-pion threshold, we can no longer apply the two-channel unitarity
and consequently the Watson theorem does not hold anymore. Therefore, the
background amplitude of the partial waves does not vanish at resonance as this
was the case for the $\Delta(1232)$ resonance. As an immediate consequence the
resonance-background separation becomes model-dependent. In MAID2007 we choose
to separate the background and resonance contributions according to the
K-matrix approximation, Eqs.~(\ref{eq:2},\ref{eq:3}). Furthermore, we recall
that the absolute values of the helicity amplitudes are correlated with the
values used for the total resonance width $\Gamma_R$ and the single-pion
branching ratio $\beta_\pi$, giving rise to additional uncertainties from these
hadronic resonance parameters. On the experimental side, the data at the higher
energies are no longer as abundant as in the $\Delta$ region. However, the
large data set recently obtained mainly by the CLAS collaboration (see
Table~\ref{tab:database}) enabled us to determine the transverse and
longitudinal helicity couplings as functions of $Q^2$ for all the four-star
resonances below 1800~MeV. These data are available in the kinematical region
of $1100~\mbox{MeV}<W<1680~\mbox{MeV}$ and
$0.4~\mbox{GeV}^2<Q^2<1.8~\mbox{GeV}^2$.

The helicity amplitudes for the Roper resonance $P_{11}(1440)$ are shown in
Fig.~\ref{fig:p11phel}. Our latest super-global solution (solid lines) is in
reasonable agreement with the single-$Q^2$ analysis. The figure shows a zero
crossing of the transverse helicity amplitude at $Q^2\approx 0.7$~GeV$^2$ and a
maximum at the relatively large momentum transfer $Q^2\approx 2.0$~GeV$^2$. The
longitudinal Roper excitation rises to large values around $Q^2\approx
0.5$~GeV$^2$ and in fact produces one of the strongest longitudinal amplitude
we can find in our analysis. This answers the question raised by Li and
Burkert\cite{Burk92} whether the Roper resonance is a radially excited 3-quark
state or a quark-gluon hybrid, because in the latter case the longitudinal
coupling should vanish completely.

\begin{center}
\includegraphics[width=3.5cm, angle=90]{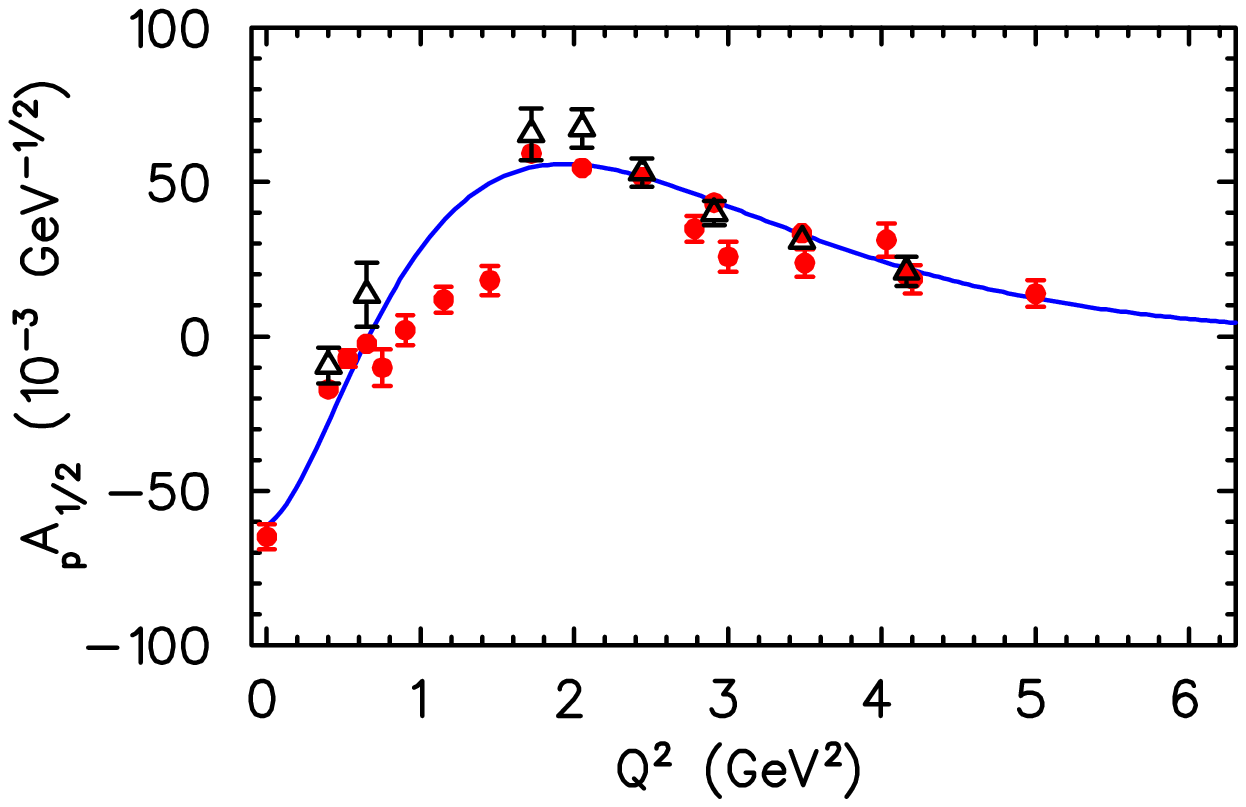}
\includegraphics[width=3.5cm, angle=90]{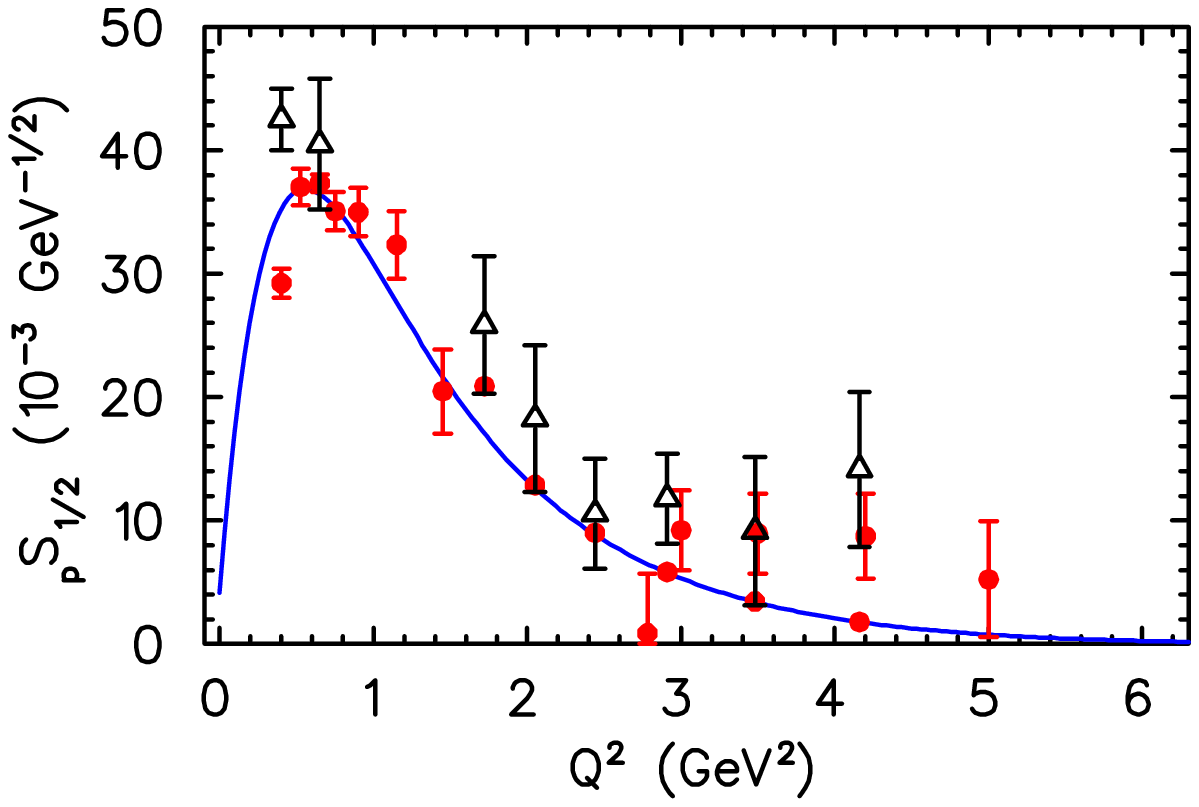}
\vspace{3mm} \figcaption{\label{fig:p11phel} Transverse and longitudinal form
factors of the $P_{11}(1440)$ Roper resonance. The red circles are the MAID
analysis of 2007\protect\cite{MAID07} and 2008, and the black triangles are the
2008 JLab analysis\protect\cite{Azn08}. The data point of the transverse form
factor at $Q^2=0$ is the PDG value\protect\cite{PDG08}.}
\end{center}

Fig.~\ref{fig:s11phel} shows our results for the $S_{11}(1535)$ resonance. The
red single-$Q^2$ data points show our 2007 analysis, while the black triangles
are the 2008 analysis of Ref.\citep{Azn08}. Our $Q^2$ dependent analysis
describes all data points quite well, except for the longitudinal form factor
in the region around $Q^2=2$~GeV$^2$.

\begin{center}
\includegraphics[width=3.5cm, angle=90]{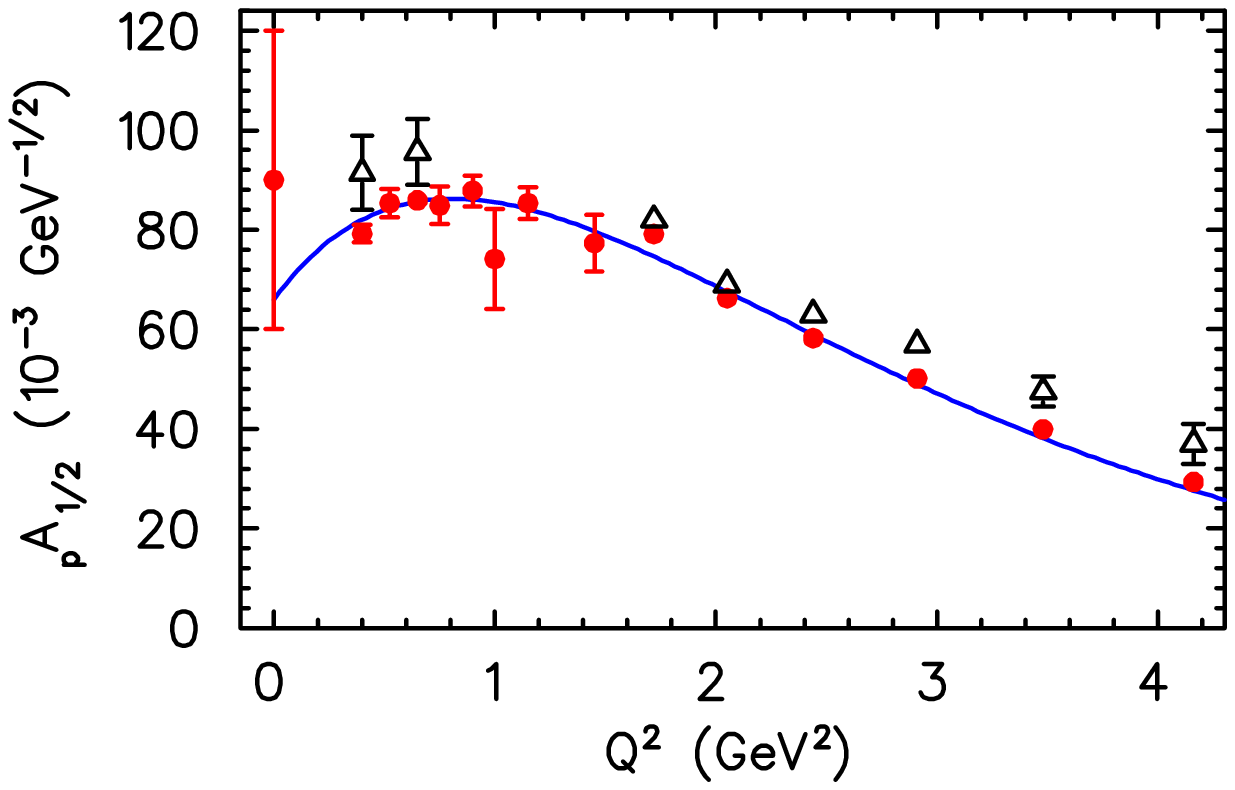}
\includegraphics[width=3.5cm, angle=90]{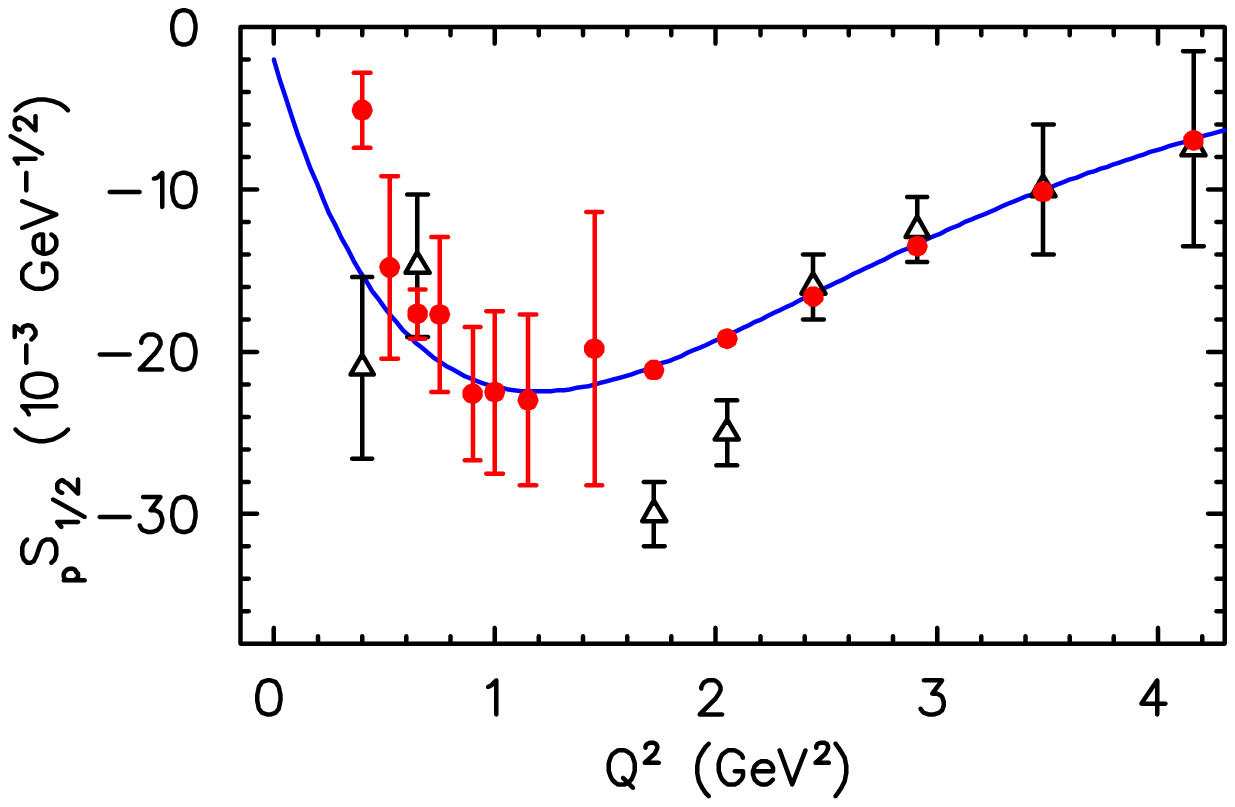}
\vspace{3mm} \figcaption{\label{fig:s11phel} Transverse and longitudinal form
factors of the $S_{11}(1535)$ resonance. Notation as in
Fig.~\ref{fig:p11phel}.}
\end{center}

While the inclusion of the 2008 Park $\pi^+$ data\cite{Par08} did not modify
our 2007 solution for the $S_{11}$ resonance, we find some significant
deviations for the $D_{13}(1520)$ resonance, see Fig.~\ref{fig:d13phel}. Also
for this resonance the JLab partial wave analysis of 2008 agrees well with the
MAID analysis for most cases, however a significant deviation remains for
$A_{1/2}$ in the region of $Q^2=2$~GeV$^2$.

\begin{center}
\includegraphics[width=3.5cm, angle=90]{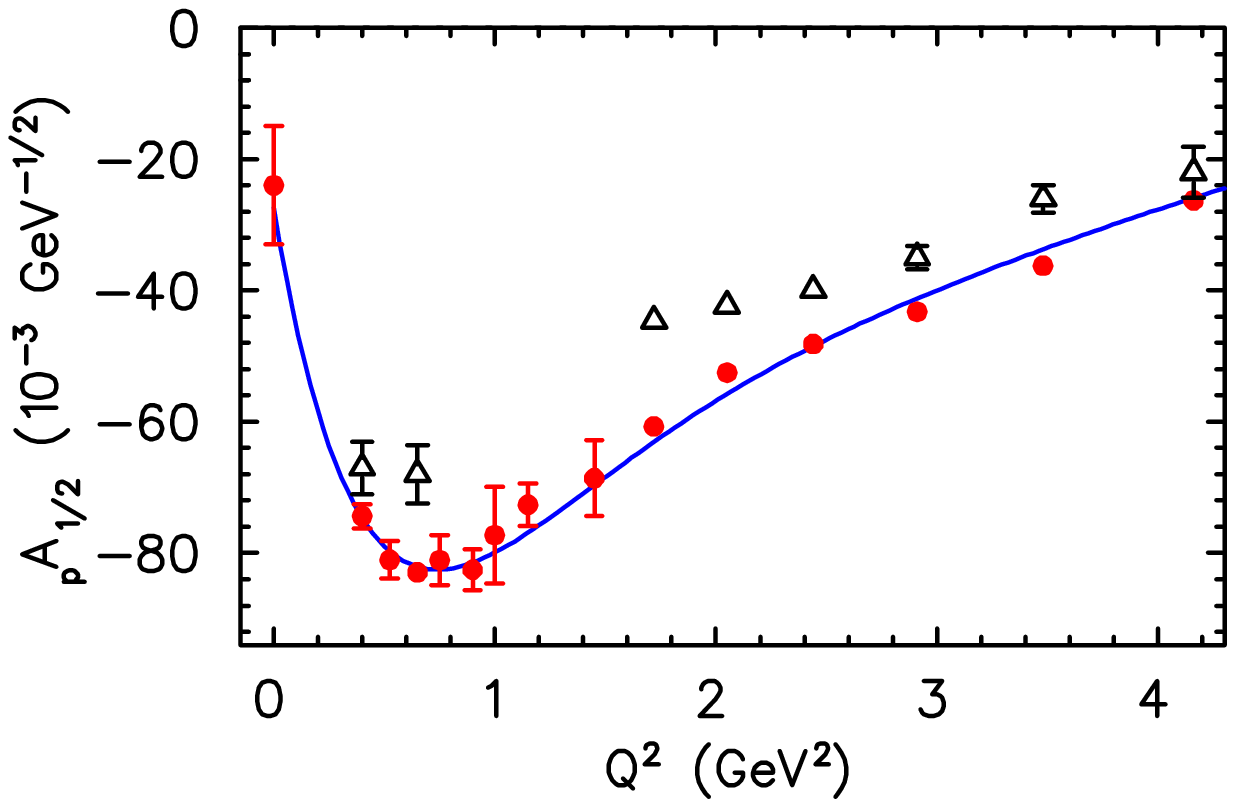}
\includegraphics[width=3.5cm, angle=90]{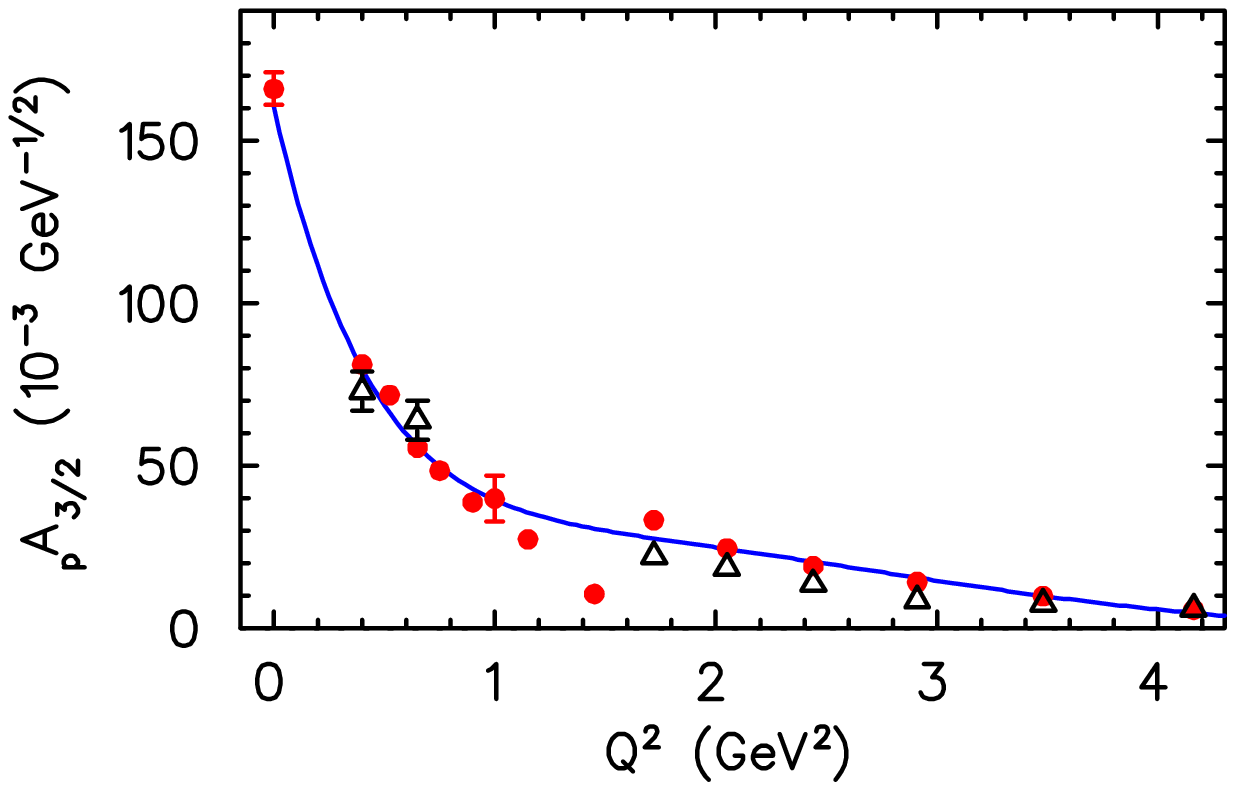}
\includegraphics[width=3.5cm, angle=90]{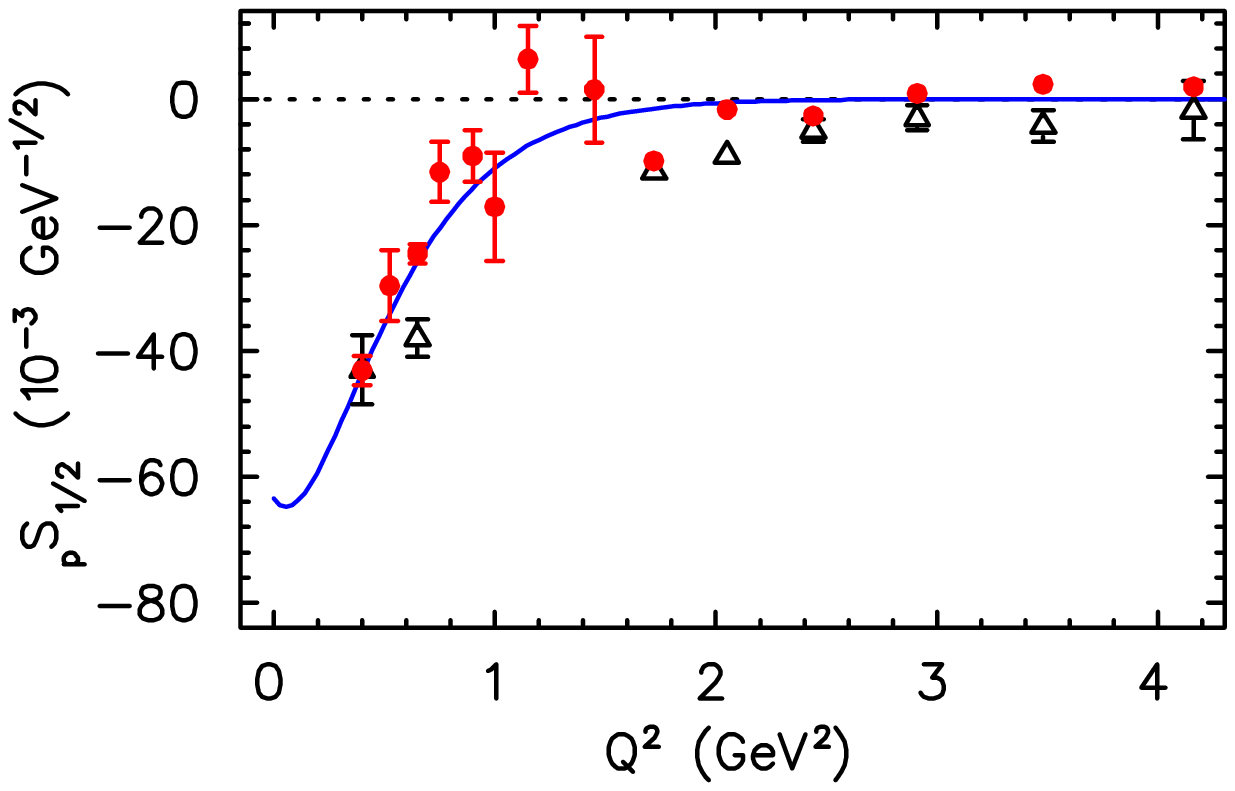}
\vspace{3mm} \figcaption{\label{fig:d13phel} Transverse and longitudinal form
factors of the $D_{13}(1520)$ resonance. Notation as in
Fig.~\ref{fig:p11phel}.}
\end{center}

\subsection{Third Resonance Region}

The $S_{31}(1620)$ is rather weakly excited by the electromagnetic probe. The
PDG $A_{1/2}$ value at the photon point is only $(27\pm 11 )\cdot 10^{-3}$
GeV$^{-1/2}$ and below $Q^2=2$~GeV$^2$ we obtain similar values, at higher
$Q^2$ it is consistent with zero, same as for the longitudinal form factor
$S_{1/2}$.
\begin{center}
\includegraphics[width=3.5cm, angle=90]{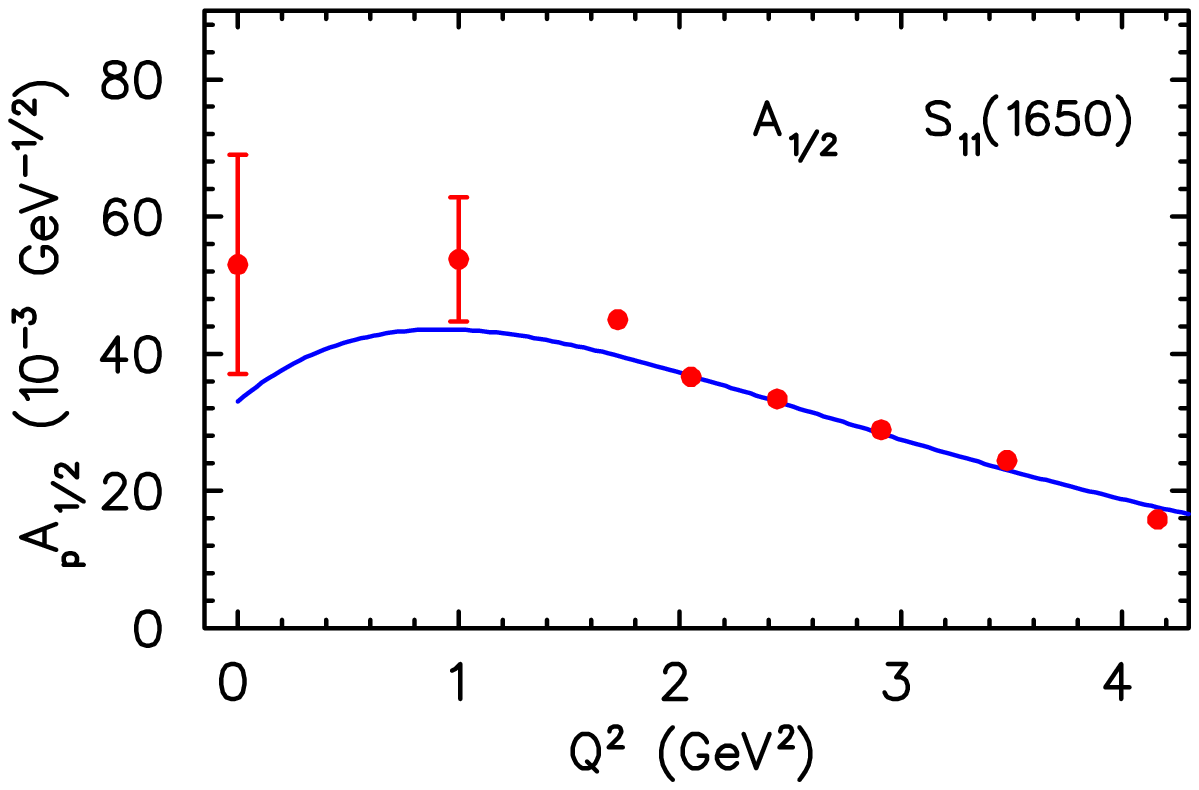}
\vspace{3mm} \figcaption{\label{fig:s11sphel} Transverse form factors of the
$S_{11}(1650)$ resonance. Notation as in Fig.~\ref{fig:p11phel}.}
\end{center}
Also for the second $S_{11}$ resonance the longitudinal coupling is practically
zero, but for the transverse form factor we find a solution shown in
Fig.~\ref{fig:s11sphel}, which has the same shape as the first $S_{11}$
resonance.

 A similar situation as for the $D_{13}$ resonance we obtain for the
$F_{15}(1680)$, shown in Fig.~\ref{fig:f15phel}.
\begin{center}
\includegraphics[width=3.5cm, angle=90]{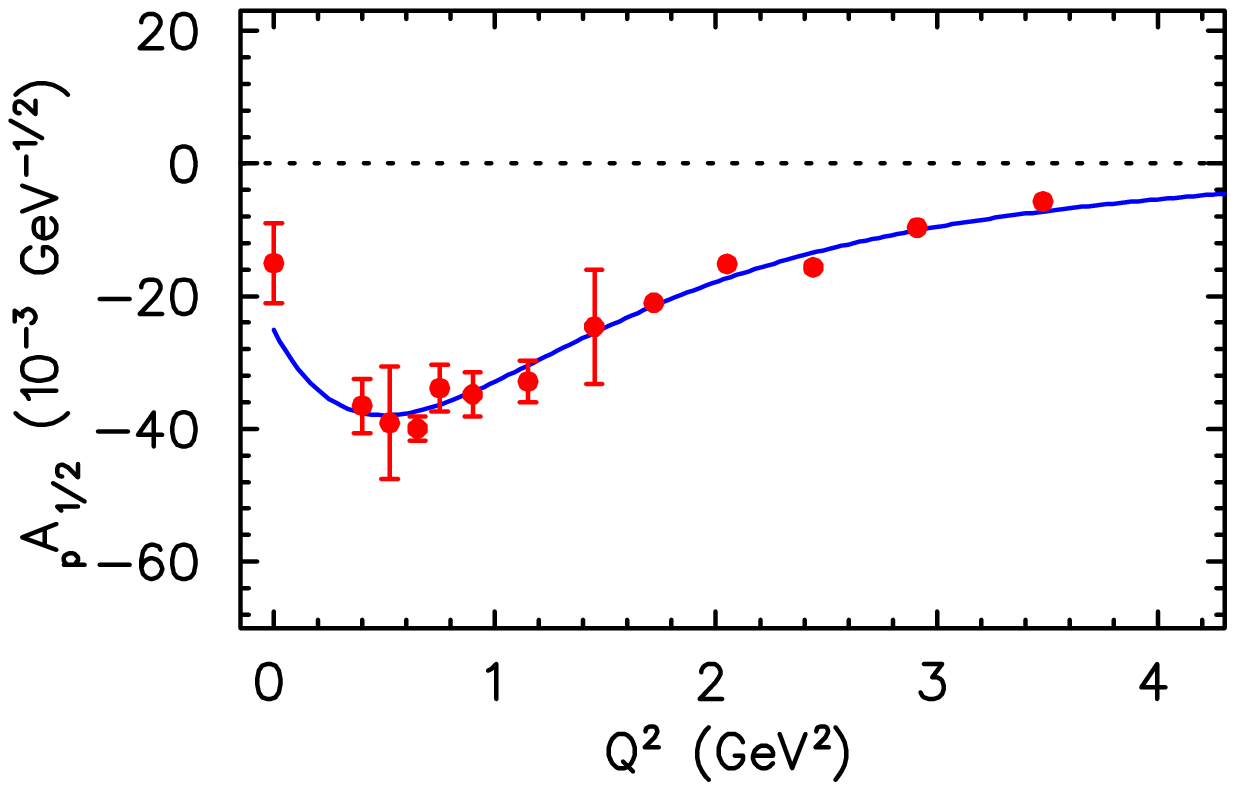}\hfill
\includegraphics[width=3.5cm, angle=90]{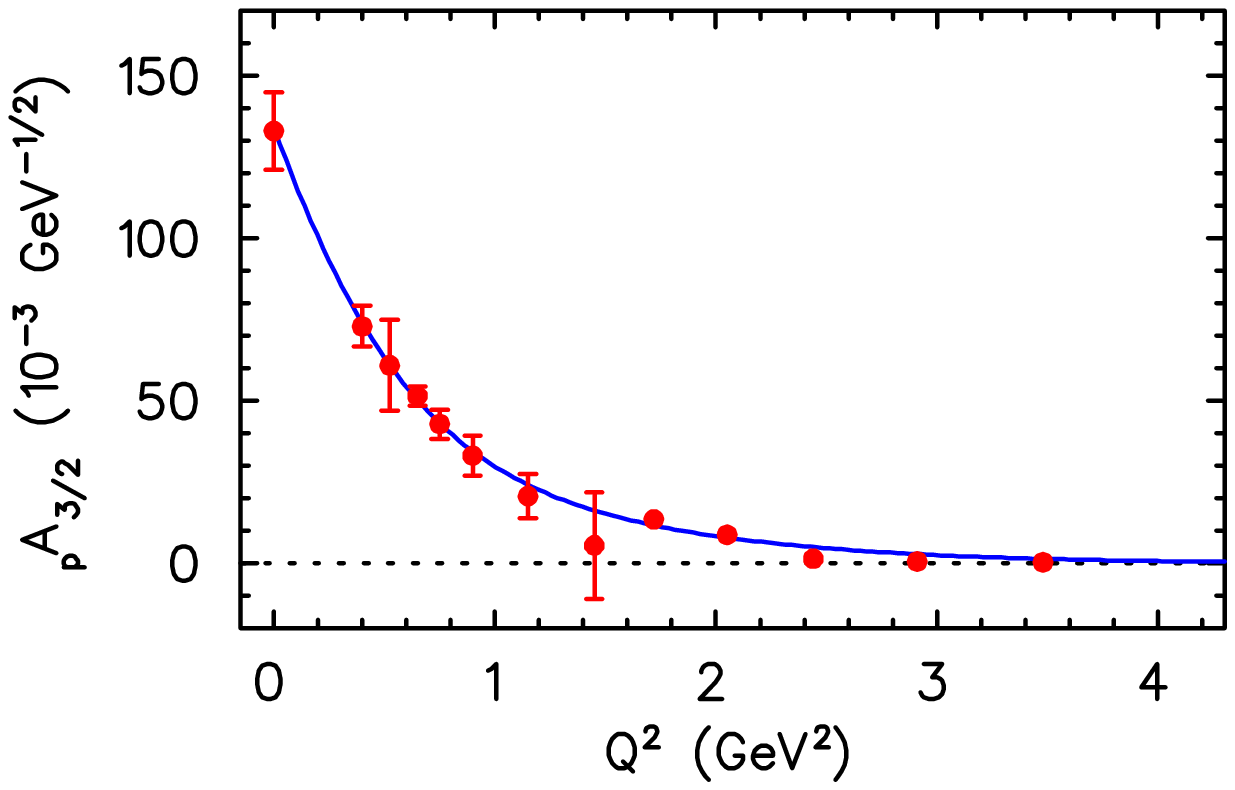}\hfill
\includegraphics[width=3.5cm, angle=90]{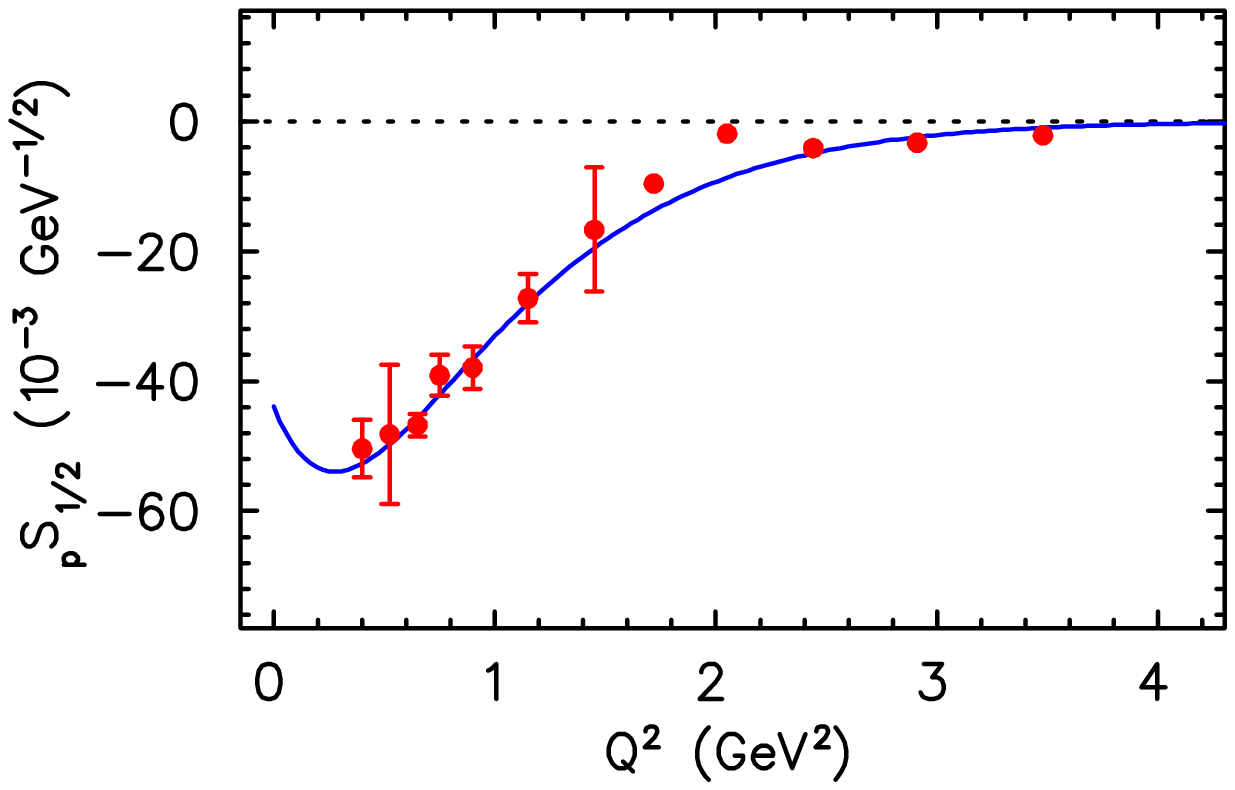}
\vspace{3mm} \figcaption{\label{fig:f15phel} Transverse and longitudinal form
factors of the $F_{15}(1680)$ resonance. Notation as in
Fig.~\ref{fig:p11phel}.}
\end{center}
For both resonances the helicity non-conserving amplitude $A_{3/2}$ dominates
for real photons and with increasing values of $Q^2$, $A_{3/2}$ drops faster
than the helicity conserving amplitude $A_{1/2}$. As a consequence the
asymmetry
\begin{equation}
{\mathcal {A}}(Q^2)=\frac{\mid A_{1/2} \mid^2 - \mid A_{3/2} \mid^2} {\mid
A_{1/2} \mid^2 + \mid A_{3/2} \mid^2} \label{eq:5.22}
\end{equation}
changes rapidly from values close to $-1$ to values near $+1$ over a small
$Q^2$ range. As a comparison, the asymmetry $\mathcal {A}$ for the
$\Delta(1232)$ resonance is practically constant over this $Q^2$ range with a
value $\approx -0.5$. This again shows the special role of the $\Delta$
resonance, where the helicity conservation is not observed.

Finally, in Figs.~\ref{fig:d15phel} and \ref{fig:p13phel} we show the situation
for the $D_{15}(1675)$ and $P_{13}(1720)$ resonances, both without a
significant longitudinal coupling. Unlike the situation discussed before, these
two resonances have dominantly helicity 3/2 transitions, whereas the $A_{1/2}$
transition is consistent with zero. As for the $\Delta(1232)$ these are further
examples for which the pQCD prediction for helicity conservation does not hold
in the $Q^2$ region below 5~GeV$^2$.
\begin{center}
\includegraphics[width=3.5cm, angle=90]{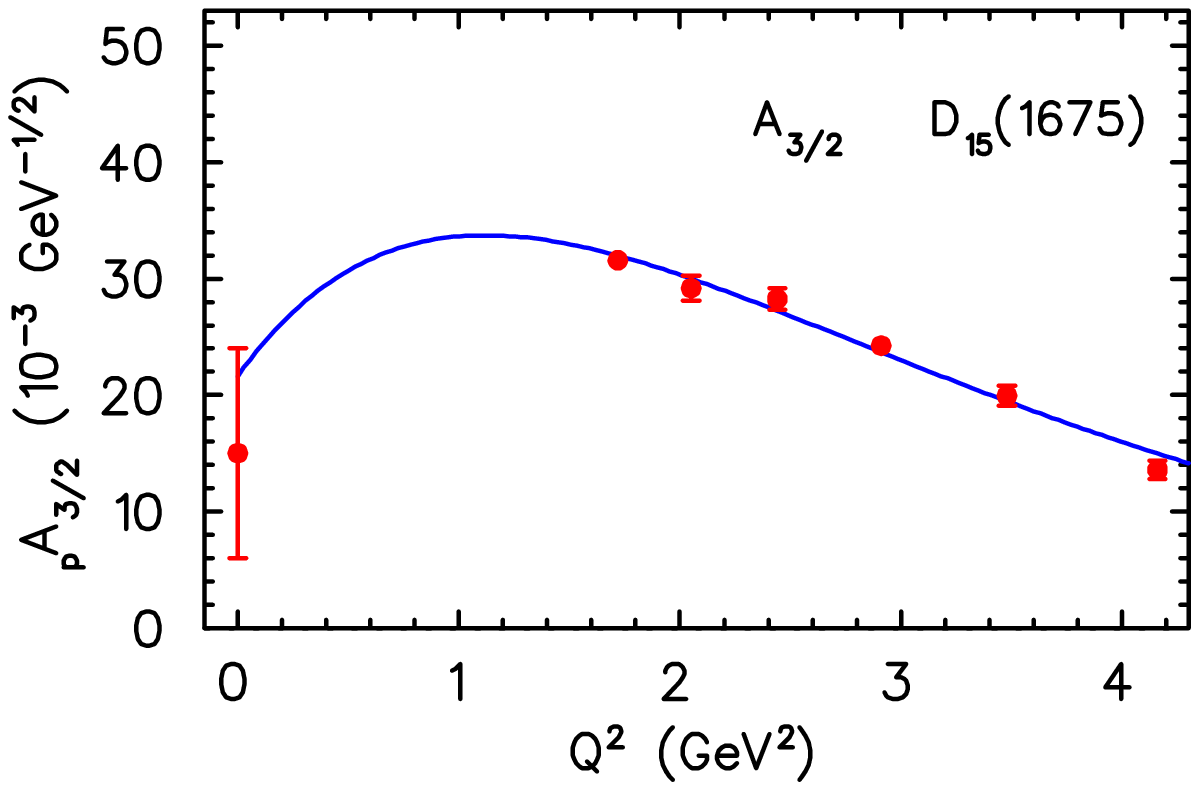}
\vspace{3mm} \figcaption{\label{fig:d15phel} Transverse form factor of the
$D_{15}(1675)$ resonance. Notation as in Fig.~\ref{fig:p11phel}.}
\end{center}
\begin{center}
\includegraphics[width=3.5cm, angle=90]{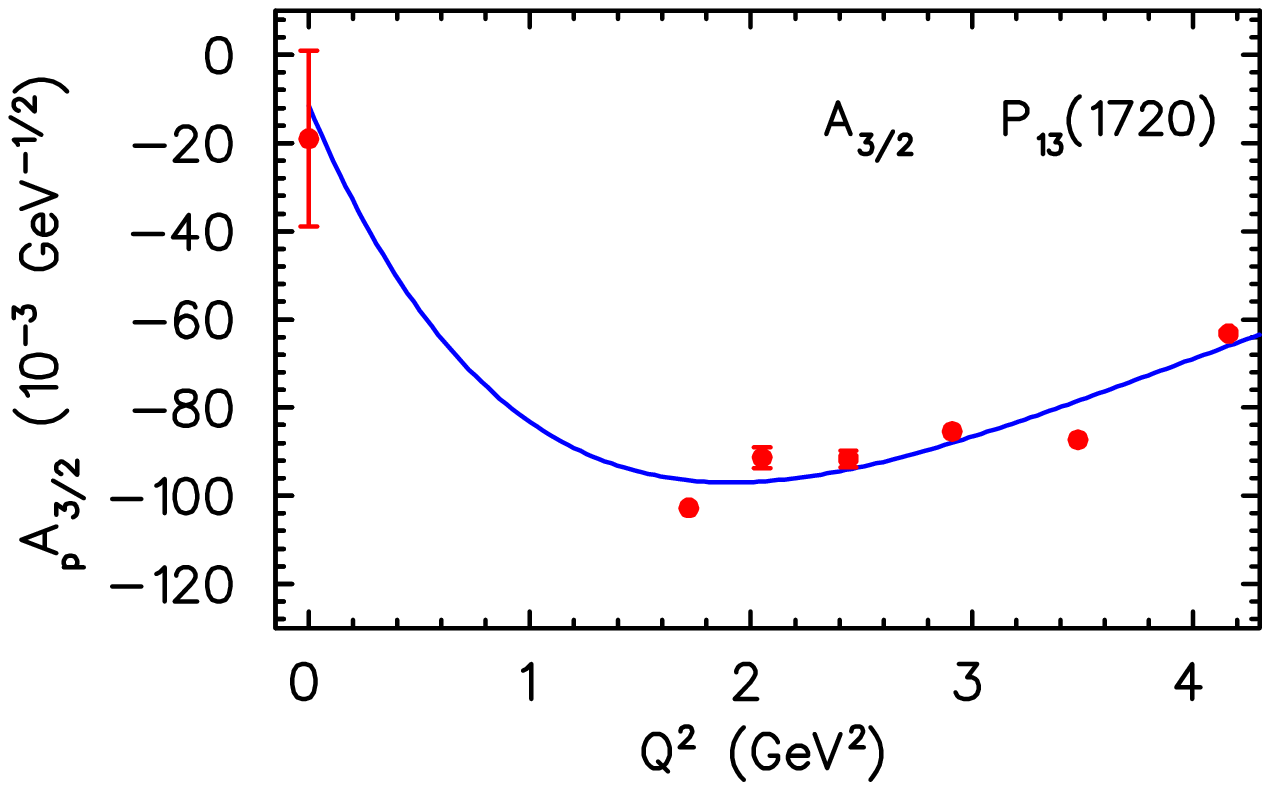}
\vspace{3mm} \figcaption{\label{fig:p13phel} Transverse form factor of the
$P_{13}(1720)$ resonance. Notation as in Fig.~\ref{fig:p11phel}.}
\end{center}

\section{Empirical transverse charge transition densities}
In the following, we will consider the e.m. $N \to N^*$ transition when viewed
from a light front moving towards the baryon. Equivalently, this corresponds
with a frame where the baryons have a large momentum-component along the
$z$-axis chosen along the direction of $P = (p + p^\prime)/2$, where $p$
($p^\prime$) are the initial (final) baryon four-momenta. We indicate the
baryon light-front + component by $P^+$ (defining $a^\pm \equiv a^0 \pm a^3$).
We can furthermore choose a symmetric frame where the virtual photon
four-momentum $q$ has $q^+ = 0$, and has a transverse component (lying in the
$xy$-plane) indicated by the transverse vector $\vec q_\perp$, satisfying $q^2
= - {\vec q_\perp}^{\, 2} \equiv - Q^2$. In such a symmetric frame, the virtual
photon only couples to forward moving partons and the + component of the
electromagnetic current $J^+$ has the interpretation of the quark charge
density operator. It is given by $J^+(0) = +2/3 \, \bar u(0) \gamma^+ u(0) -
1/3 \, \bar d(0) \gamma^+ d(0)$, considering only $u$ and $d$ quarks. Each term
in the expression is a positive operator since $\bar q \gamma^+ q \propto |
\gamma^+ q |^2$.

We define a transition charge density for the unpolarized $N \to N^\ast$
transition by the Fourier transform
\begin{eqnarray}
\rho_0^{N N^\ast}(\vec b) &\equiv& \int \frac{d^2 \vec q_\perp}{(2 \pi)^2} \,
e^{- i \, \vec q_\perp \cdot \vec b} \, \frac{1}{2 P^+}\times\nonumber\\
&&\langle P^+, \frac{\vec q_\perp}{2}, \lambda \,|\, J^+(0) \,|\,
P^+, -\frac{\vec q_\perp}{2}, \lambda  \rangle ,\hspace*{0.7cm}
\label{eq:ndens0}
\end{eqnarray}
where $\lambda$ denotes the nucleon and $N^\ast$ light-front helicities, $\vec
q_\perp = Q ( \cos \phi_q \hat e_x + \sin \phi_q \hat e_y )$, and where the
2-dimensional vector $\vec b$ denotes the position (in the $xy$-plane) from the
transverse {\it c.m.} of the baryons.

First we will consider the case of $j=1/2$ resonances, as $P_{11}$ and
$S_{11}$. These cases are very similar to the nucleon and can be worked out in
an analogous way. The Fourier transform in Eq.~(\ref{eq:ndens0}) leads to
\begin{eqnarray}
\rho_0^{N N^\ast}(\vec b) =
\int_0^\infty \frac{d Q}{2 \pi} Q \, J_0(b \, Q) F_1^{N N^\ast}(Q^2),
\label{eq:ndens1}
\end{eqnarray}
where $J_n$ denotes the cylindrical Bessel function of order $n$.
Note that $\rho_0^{N N^\ast}$ only depends on $b = |\vec b|$.
It has the interpretation of the quark (transition) charge density in the
transverse plane which induces the $N \to N^\ast$ excitation.

The above unpolarized transition charge density involves only one of the two
independent $N \to N^\ast$ e.m. form factors. To extract the information
encoded in $F_2^{N N^\ast}$, we consider the transition charge densities for a
transversely polarized $N$ and $N^\ast$. We denote this transverse polarization
direction by $\vec S_\perp = \cos \phi_S \hat e_x + \sin \phi_S \hat e_y$. The
transverse spin state can be expressed in terms of the light front helicity
spinor states as $| s_\perp = + \frac{1}{2} \rangle = \left( | \lambda = +
\frac{1}{2} \rangle + e^{i \phi_S } \, | \lambda = - \frac{1}{2} \rangle
\right) / \sqrt{2}$, with $s_\perp$ the nucleon spin projection along the
direction of $\vec S_\perp$.

We can then define a transition charge density for a transversely polarized $N$
and $N^\ast$, both along the direction of $\vec S_\perp$ as
\begin{eqnarray}
\rho_T^{N N^\ast}(\vec b) &\equiv& \int \frac{d^2 \vec q_\perp}{(2 \pi)^2} \,
e^{-i \, \vec q_\perp \cdot \vec b} \, \frac{1}{2 P^+}\times\nonumber\\
&&\langle P^+, \frac{\vec q_\perp}{2}, s'_\perp \,|\, J^+(0) \,|\, P^+,
-\frac{\vec q_\perp}{2}, s_\perp \rangle .\hspace*{0.7cm} \label{eq:ndens2}
\end{eqnarray}
Using Eq.~(\ref{eq:ndens1}), the Fourier transform of Eq.~(\ref{eq:ndens2}) can
be worked out for the case $s'_\perp=s_\perp$ as
\begin{eqnarray}
\lefteqn{\rho_T^{N N^\ast}(\vec b) = \rho_0^{ N N^\ast}(b)
+ \sin (\phi_b - \phi_S) \,\times\hspace{1.2cm}}  \nonumber\\
&&\int_0^\infty \frac{d Q}{2 \pi} \frac{Q^2}{(M^\ast + M_N)} \, J_1(b \, Q)
F_2^{N N^\ast}(Q^2), \label{eq:ndens3}
\end{eqnarray}
where the second term, which describes the deviation from the circular
symmetric unpolarized charge density, depends on the orientation of $\vec b = b
( \cos \phi_b \hat e_x + \sin \phi_b \hat e_y )$. In the following we choose
the transverse spin along the $x$-axis ($\phi_S = 0$).

\begin{center}
\includegraphics[width =3.8cm]{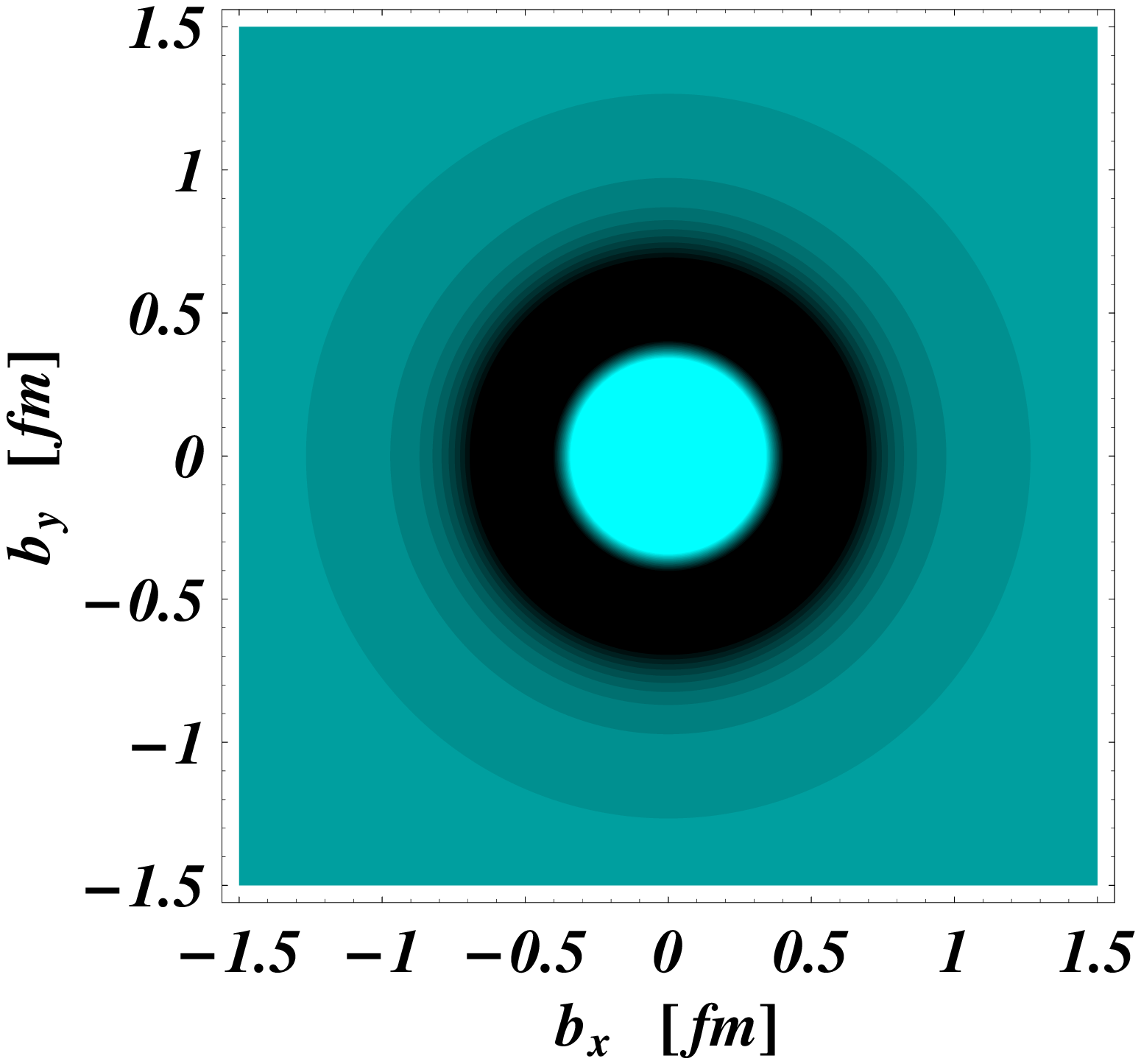}
\includegraphics[width =3.8cm]{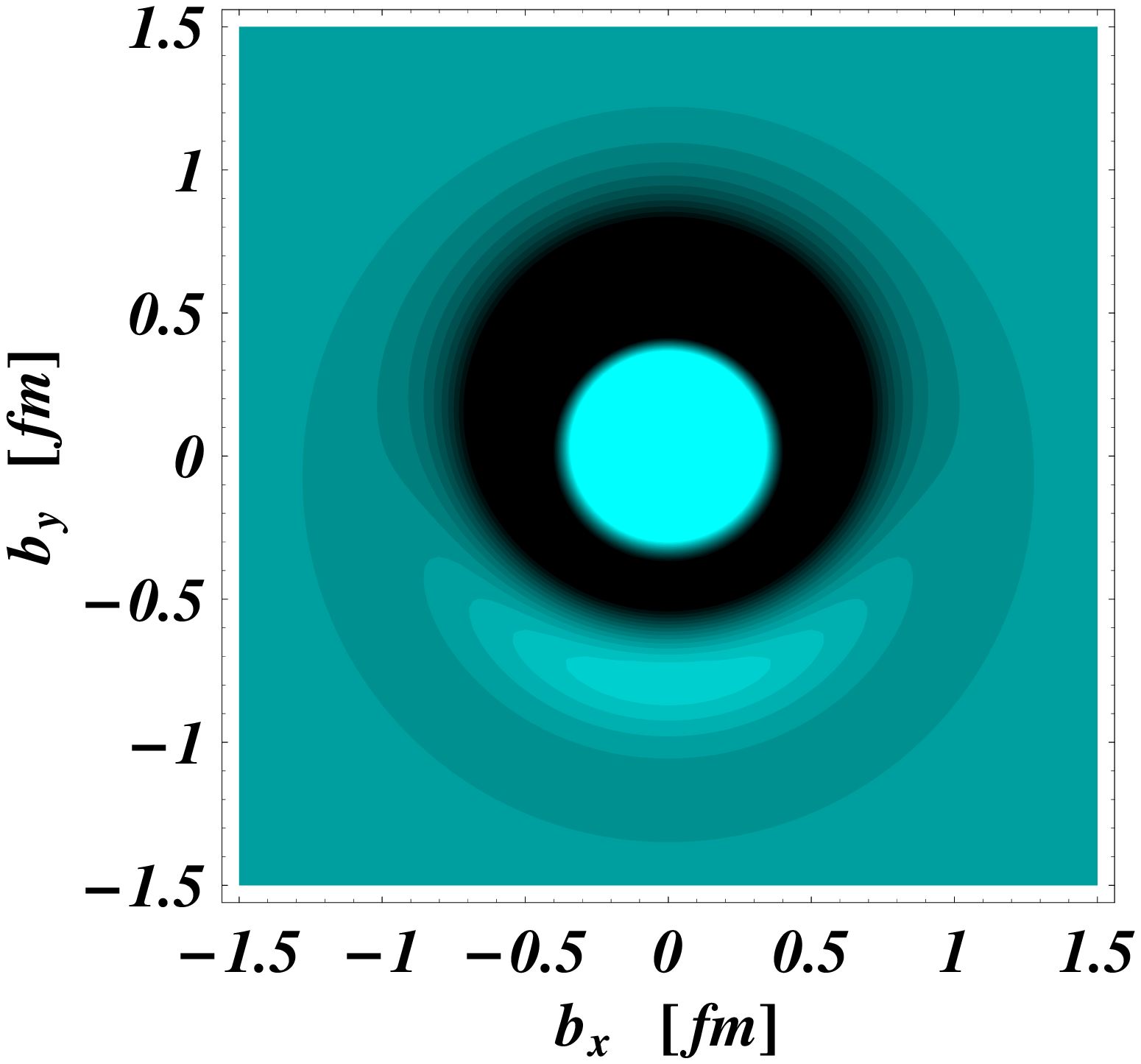}
\vspace{3mm} \figcaption{\label{fig:p11dens1p} Quark transverse charge density
corresponding to the $p \to P_{11}(1440)$ e.m. transition. Left panel: When $p$
and $N^\ast$ are unpolarized ($\rho_0^{p N^\ast}$). Right panel: When $p$ and
$N^\ast$ are polarized along the $x$-axis ($\rho_T^{p N^\ast}$). The light
(dark) regions correspond with positive (negative) densities. For the $p \to
P_{11}(1440)$ e.m. transition FFs, we use the improved MAID2008 fit of this
work.}
\end{center}

In Fig.~\ref{fig:p11dens1p} we show the results for the $N \to P_{11}(1440)$
transition charge densities both for the unpolarized case and for the case of
transverse polarization in for the proton and in Fig.~\ref{fig:p11dens1n} for
the neutron. We use the empirical information on the $N \to N^\ast(1440)$
transition FFs as given in our parametrization of this work.

It is seen that for the transition on a proton, which is well constrained by
data, there is an inner region of positive quark charge concentrated within
0.5~fm, accompanied by a relatively broad band of negative charge extending out
to about 1~fm.
\begin{center}
\includegraphics[width =3.8cm]{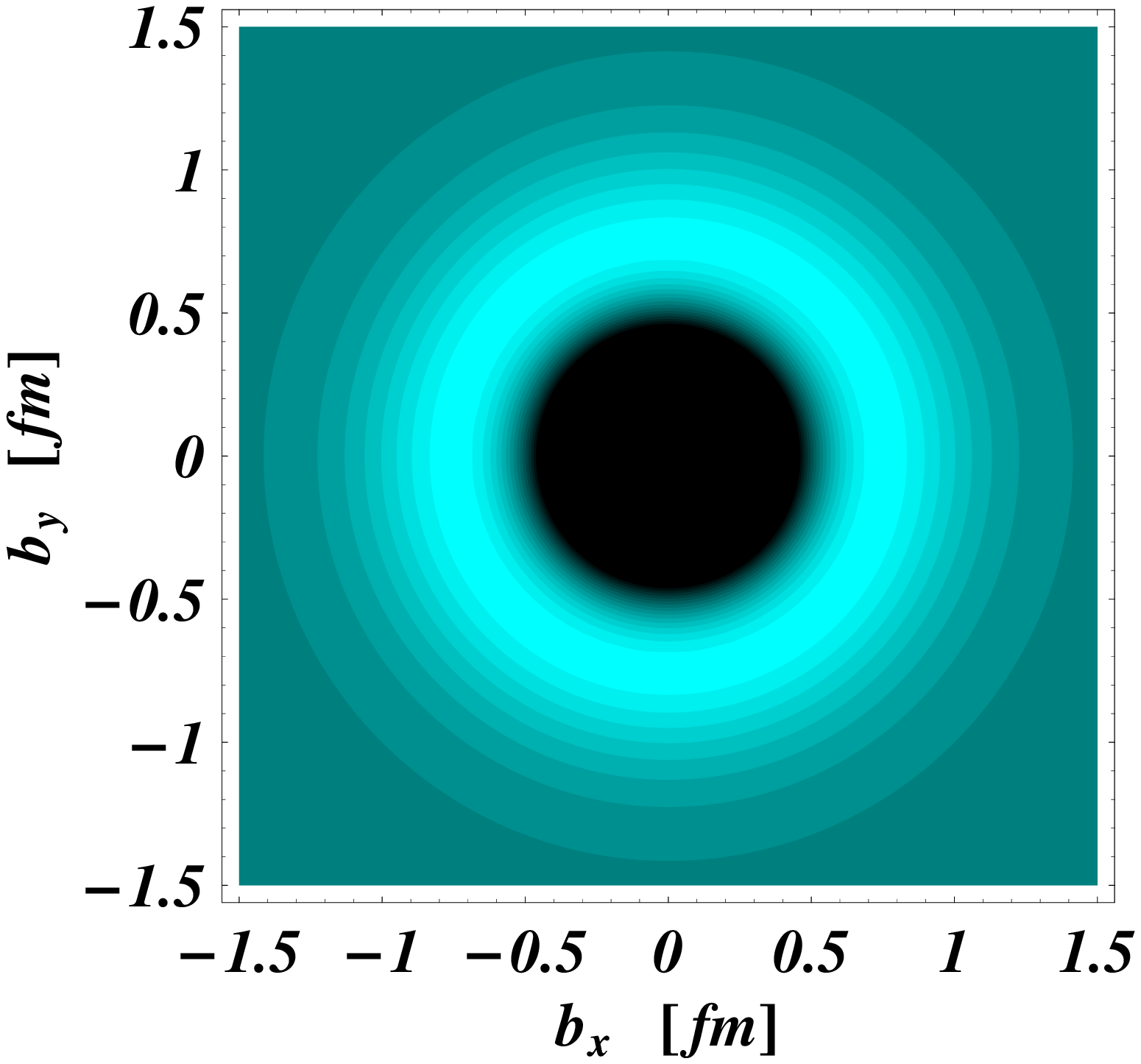}
\includegraphics[width =3.8cm]{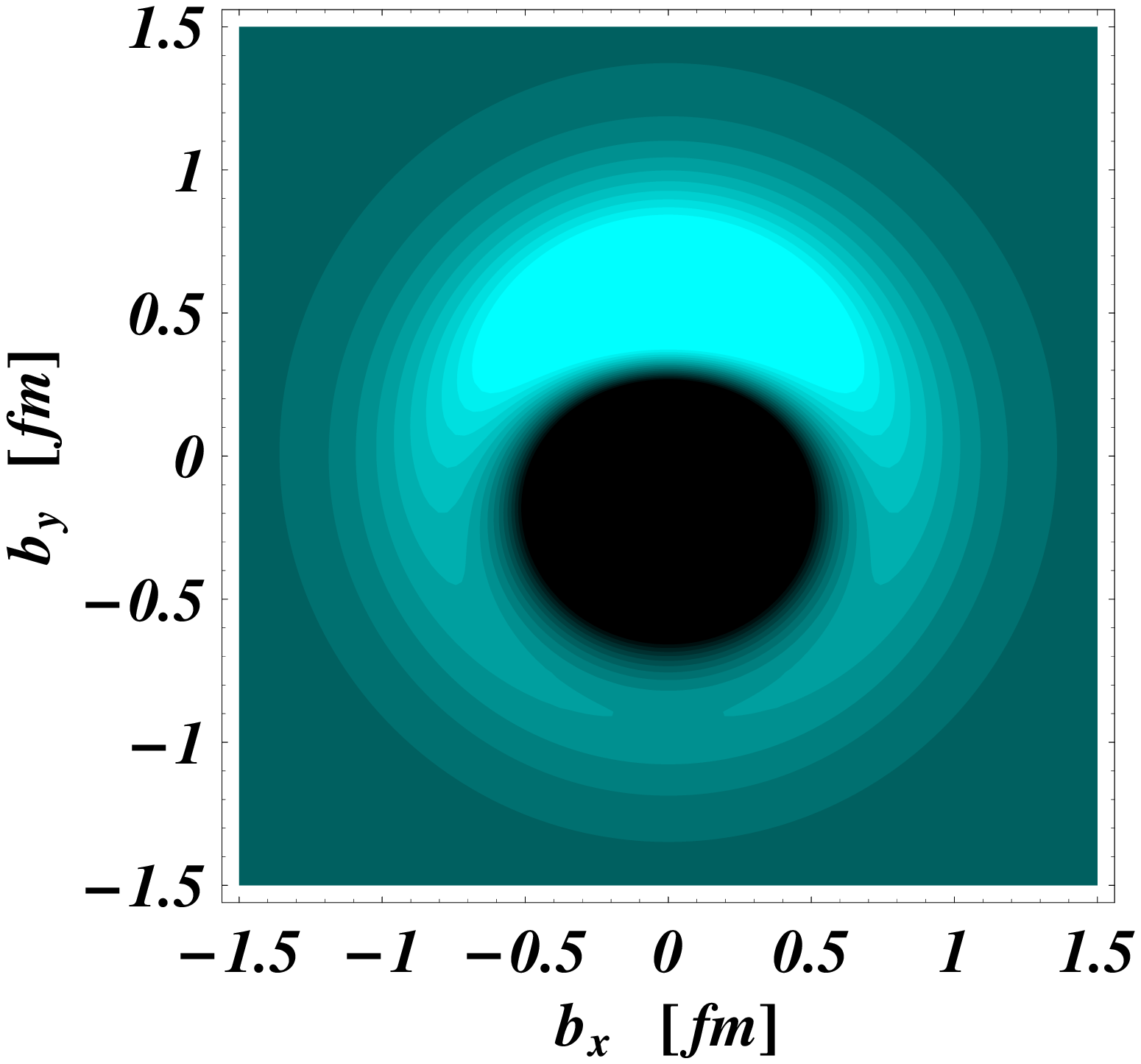}
\vspace{3mm} \figcaption{\label{fig:p11dens1n}Quark transverse charge density
corresponding to the $n \to P_{11}(1440)$ e.m. transition. Left panel: When $n$
and $N^\ast$ are unpolarized ($\rho_0^{n N^\ast}$). Right panel: When $n$ and
$N^\ast$ are polarized along the $x$-axis ($\rho_T^{n N^\ast}$). The light
(dark) regions correspond with positive (negative) densities. For the $n \to
P_{11}(1440)$ e.m. transition FFs, we use the MAID2007 fit.}
\end{center}
When polarizing the baryon in the transverse plane, the large value of the
magnetic transition strength at the real photon point, yields a sizeable shift
of the charge distribution, inducing an electric dipole moment. For the
neutron, which is not very well constrained by data, the MAID2007 analysis
yields charge distributions of opposite sign compared to the proton, with
active quarks spreading out over even larger spatial distances.

Fig.~\ref{fig:s11dens1p} shows the unpolarized and polarized transition charge
densities from the proton to the $S_{11}(1535)$ resonance. It can be compared
to the corresponding Fig.~\ref{fig:p11dens1p} for the Roper. The up quarks are
not so strongly localized and also the ring of down quarks is less pronounced,
in particular for the density corresponding to $F_2^{NN^*}$ in the second
panel.
\begin{center}
\includegraphics[width=3.6cm]{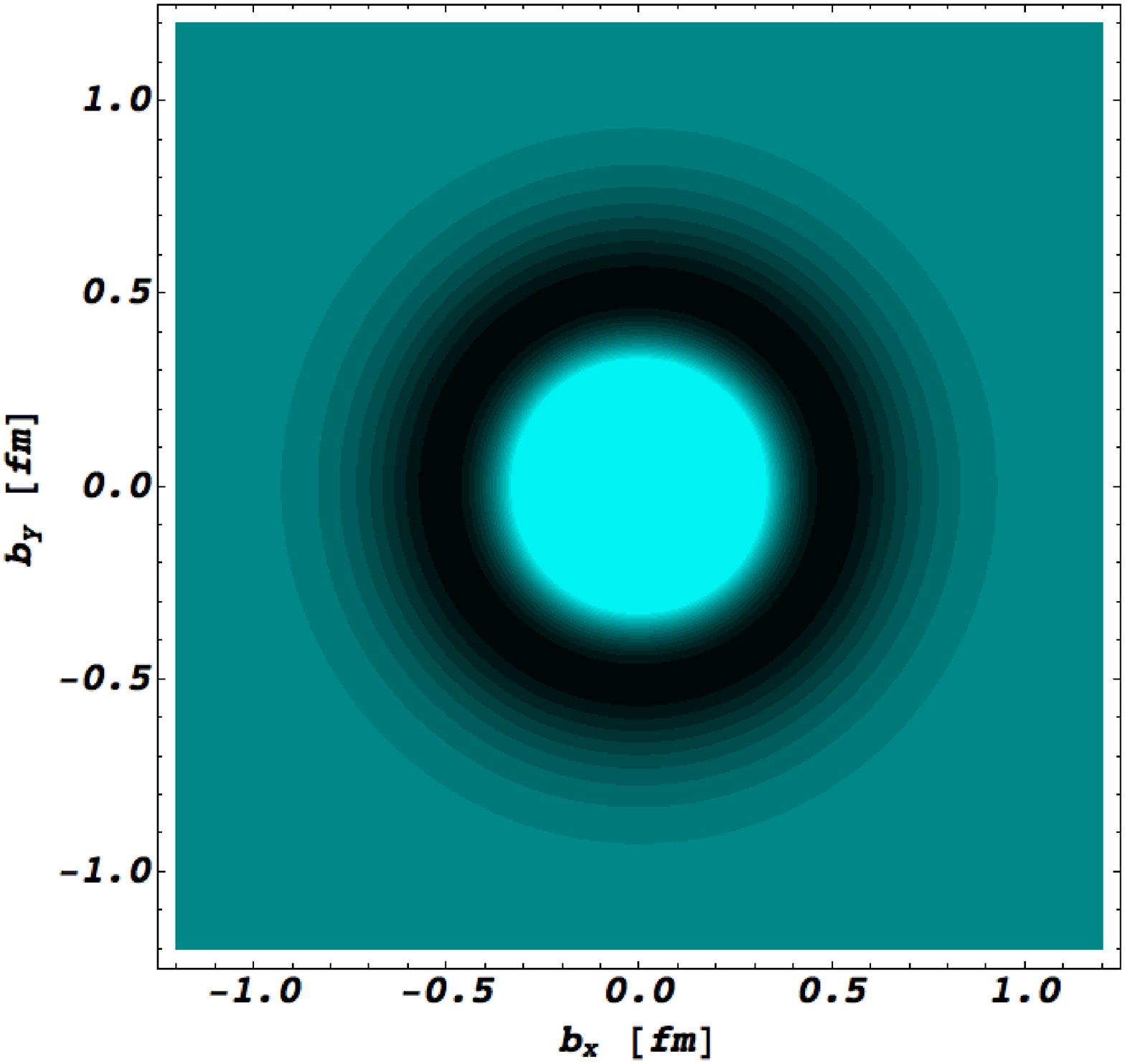}
\hspace{0.0cm}
\includegraphics[width=3.6cm]{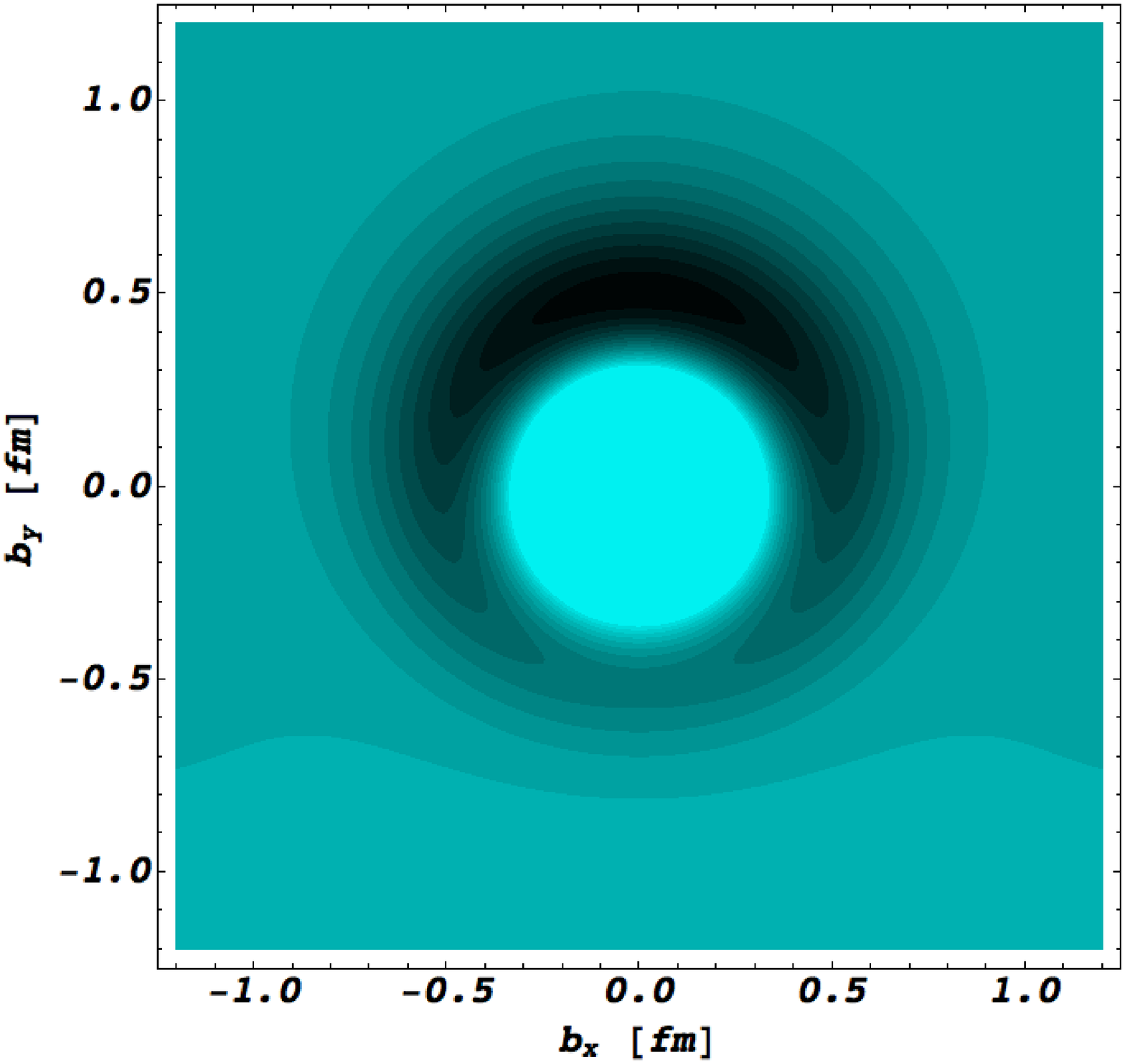}
\vspace{3mm} \figcaption{\label{fig:s11dens1p} Quark transverse charge density
corresponding to the $p \to S_{11}(1535)$ e.m. transition. Left panel: When $p$
and $S_{11}$ are in a light-front helicity +1/2 state ($\rho_0^{p S_{11}}$).
Right panel: When $p$ and $S_{11}$ are polarized along the $x$-axis with
opposite spin projections ($\rho_T^{p S_{11}}$), i.e. $s_\perp=-s'_\perp=+1/2$.
The light (dark) regions correspond with positive (negative) densities. For the
$p \to S_{11}(1535)$ e.m. transition FFs, we use the MAID2007 fit. }
\end{center}

Finally, in Fig.~\ref{fig:d13dens1p} we show two transition densities from the
proton to the $D_{13}(1520)$ resonance. Similar to the $N\Delta(1232)$
transition, also here we have 3 FFs leading to 3 densities.
\begin{center}
\includegraphics[width=3.6cm]{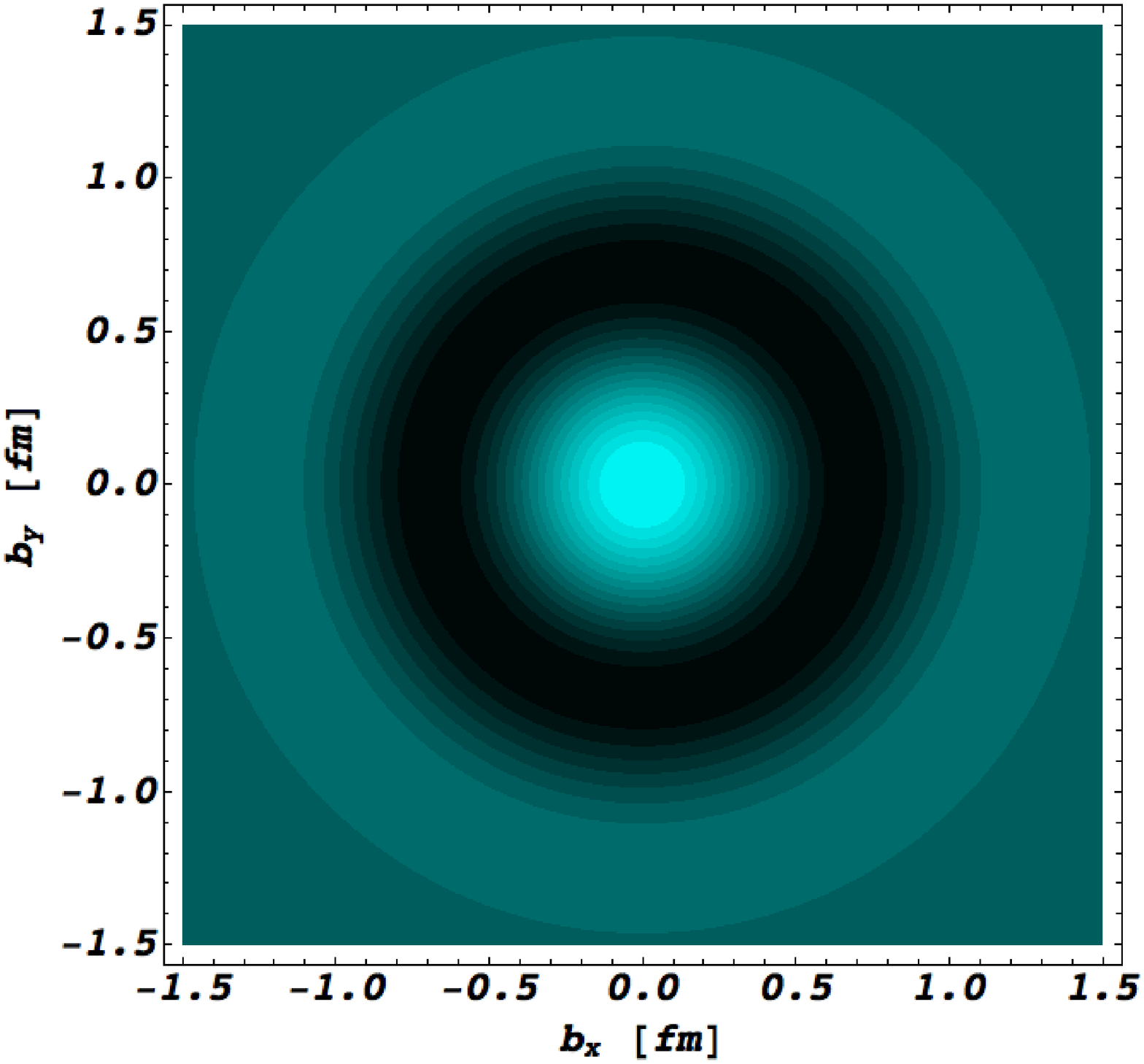}
\hspace{0.0cm}
\includegraphics[width=3.6cm]{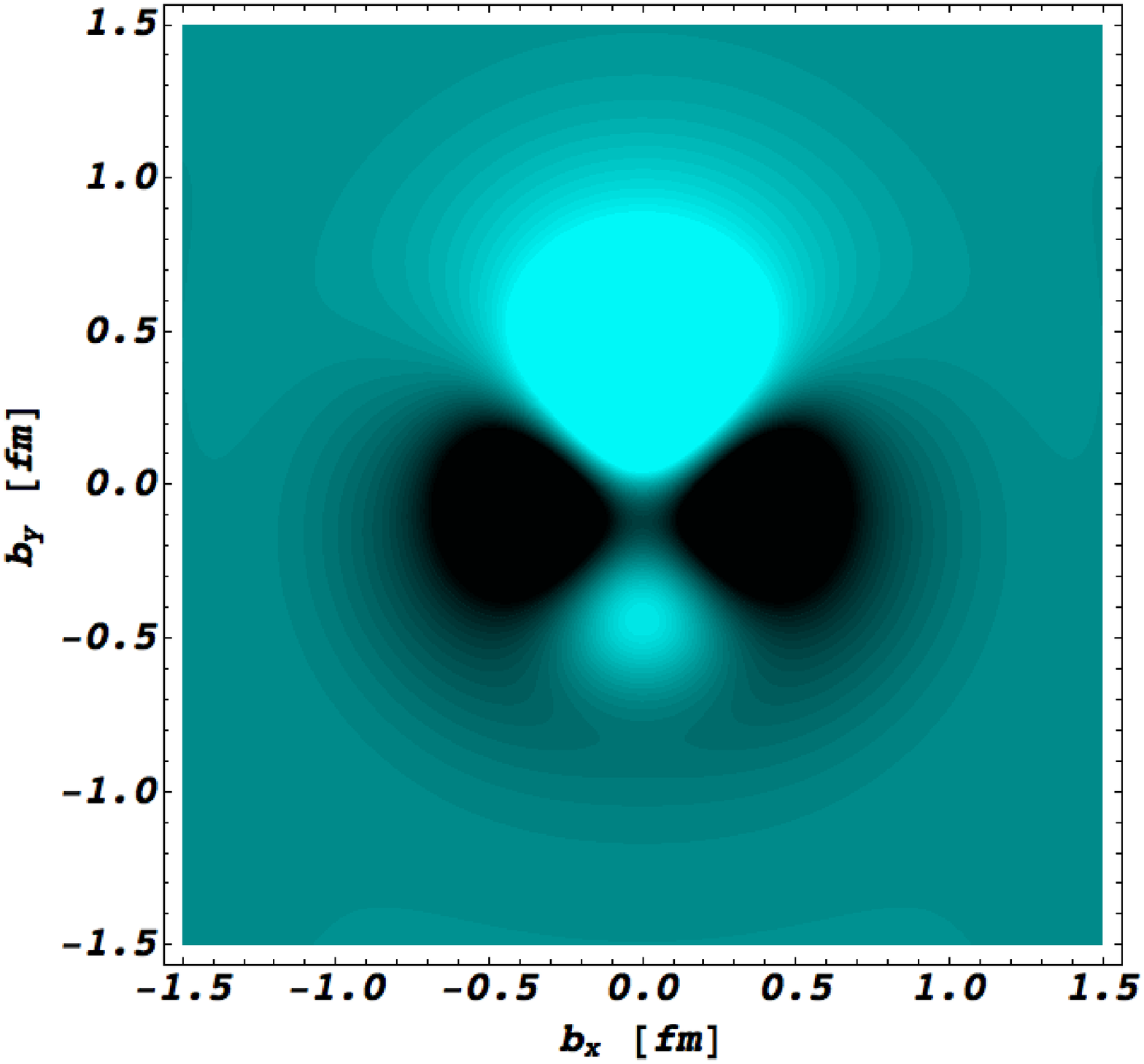}
\vspace{3mm} \figcaption{\label{fig:d13dens1p} Quark transverse charge density
corresponding to the $p \to D_{13}(1520)$ e.m. transition. Left panel: When $p$
and $D_{13}$ are in a light-front helicity +1/2 state ($\rho_0^{p D_{13}}$).
Right panel: When $p$ and $D_{13}$ are polarized along the $x$-axis with spin
projections ($\rho_T^{p D_{13}}$) as in Fig.~\ref{fig:s11dens1p}. The light
(dark) regions correspond with positive (negative) densities. For the $p \to
D_{13}(1520)$ e.m. transition FFs, we use the improved MAID2008 fit of this
work. }
\end{center}
The unpolarized density is similar to the Roper with, however, more diffuse
boundaries between up and down quarks. In addition to the dipole transition
density, in this case we also get a quadrupole density which is shown in the
second panel.

\section{Summary and Conclusions}
Using the world data base of pion photo- and electroproduction and recent data
from Mainz, Bonn, Bates and JLab we have made a first attempt to extract all
longitudinal and transverse helicity amplitudes of nucleon resonance excitation
for four-star resonances below $W=2$~GeV. For this purpose we have extended our
unitary isobar model MAID and have parameterized the $Q^2$ dependence of the
transition amplitudes. Comparisons between single-$Q^2$ fits and a $Q^2$
dependent superglobal fit give us confidence in the determination of the
helicity couplings of the $P_{33}(1232), P_{11}(1440), S_{11}(1535),
D_{13}(1520)$ and the $F_{15}(1680)$ resonances, even though the model
uncertainty of these amplitudes can be as large as 50\% for the longitudinal
amplitudes of the $D_{13}$ and $F_{15}$.

These form factors were used to extract the quark transverse charge densities
inducing these transitions. The rings of up and down quarks in these
two-dimensional representations show very different structures for the Roper,
the $S_{11}$ and the $D_{13}$ resonances.

\end{multicols}

\end{document}